\documentclass[aps,prd,preprint,nofootinbib]{revtex4}
\usepackage[subpreambles=true]{standalone}
\usepackage{import}
\usepackage{amsfonts}
\usepackage{mathrsfs}
\usepackage{graphicx}
\usepackage{appendix}
\usepackage{amsmath}
\usepackage{amssymb}
\usepackage{epsfig}
\usepackage{graphicx}
\usepackage{slashed}
\usepackage{color}
\usepackage{multirow}
\usepackage{makecell}                                      
\usepackage{float}
\usepackage{ulem}
\usepackage{adjustbox}
\usepackage{feynmp}
\usepackage{tikz-feynman}
\tikzfeynmanset{compat=1.1.0}

\usepackage{stmaryrd}
\allowdisplaybreaks[1]

\graphicspath{{fig/}{dia/}} \DeclareGraphicsExtensions{.eps}
\usepackage[colorlinks,
            linkcolor=blue,
            anchorcolor=blue,
            citecolor=blue]{hyperref}
\begin{document}

\title{	Nonleptonic three-body charmed baryon weak decays with $H ( {\bf 15}) $ }

\author {Chao-Qiang Geng$^1$, 
Chia-Wei Liu$^{2,3}$ and 	
Sheng-Lin Liu$^{1,4,5}$\footnote{liushenglin22@mails.ucas.ac.cn}}
\affiliation{
$^1$School of Fundamental Physics and Mathematical Sciences, Hangzhou Institute for Advanced Study,~UCAS, Hangzhou 310024, China\\
$^2$Tsung-Dao Lee Institute,
Shanghai Jiao Tong University, Shanghai 200240, China\\
$^3$Key Laboratory for Particle Astrophysics and Cosmology (MOE)
\& Shanghai Key Laboratory for Particle Physics and Cosmology,
Shanghai Jiao Tong University, Shanghai 200240, China\\
$^4$Institute of Theoretical Physics,~UCAS, Beijing 100190, China\\
$^5$University of Chinese Academy of Sciences, 100190 Beijing, China
}
\date{\today}

\begin{abstract}
We study the nonleptonic three-body charmed baryon weak decays of $\mathbf{B}_{c}\rightarrow\mathbf{B}_{n}PP^{\prime}$ under the $SU(3)_{F}$ flavor symmetry, where $\mathbf{B}_{c}$ denotes the anti-triplet charmed baryon, comprising $(\Xi^{0}_{c},-\Xi^{+}_{c},\Lambda^{+}_{c})$, and $\mathbf{B}_{n}$ and $P(P^{\prime})$ represent octet baryon and pseudoscalar meson states, respectively. 
In addition to 12 parameters from the contributions of the color-antisymmetric part of the effective Hamiltonian, denoted as $H(\bar{\mathbf{6}})$, there are 4 parameters from the color-symmetric one, $H(\mathbf{15})$, which were not included in the previous study. With 16 parameters in total and 
28 experimental data points, we obtain the minimal $\chi^2$ over degree of freedom of  $\chi^{2}/d.o.f=1.5$, which is a great improvement comparing to that without $H(\mathbf{15})$. With the better fitting values, we evaluate the branching ratios and up-down asymmetries of $\mathbf{B}_{c}\rightarrow\mathbf{B}_{n}PP^{\prime}$, which present some interesting results such as $\mathcal{B}\,(\Lambda^{+}_{c}\rightarrow(\Xi(1690)^{0}\rightarrow\Sigma^{+}K^{-})\,K^{+})\equiv(1.5\pm0.4)\times10^{-3}$ and potential $SU(3)$ breaking effects in $\Xi^{+}_{c}\rightarrow p\pi^{+}K^{-}$ and $\Lambda^{+}_{c}\rightarrow \Sigma^{+}\pi^{-}K^{+}$ to be verified by the experiments at BESIII, Belle-II and LHCb.

\end{abstract}

\maketitle

\section{Introduction}\label{Introduction}

The nonleptonic three-body charmed baryon weak decays of $\mathbf{B}_{c}\rightarrow\mathbf{B}_{n}PP^{\prime}$ have been searched continuously by Belle-II~\cite{Belle:2018kzz,Belle:2019bgi,Belle:2020xku,Belle:2022pwd,Belle:2020ito}, BESIII~\cite{BESIII:2018qyg,BESIII:2023rky,BESIII:2020kzc,BESIII:2023dvx,BESIII:2023pia,BESIII:2023sdr,BESIII:2018cvs} and LHCb~\cite{LHCb:2017xtf,LHCb:2020gge} Collaborations with increasing high precision, where $\mathbf{B}_{c}\equiv(\Xi^{0}_{c},-\Xi^{+}_{c},\Lambda^{+}_{c})$ represents the anti-triplet charmed baryon, while $\mathbf{B}_{n}$ and $P(P^{\prime})$ denote octet baryon and pseudoscalar meson states, respectively. A systematic study of these three-body charmed baryon decays is crucial due to their rich spin structures, providing insights into P, CP, or T-violating spin correlations, and aiding in understanding the complete dynamics of baryonic decay processes.

In contrast to the wealth of observational data, calculating charm quark three-body weak decays into light quarks has proven challenging. This is attributed to the large mass of the charm quark, making the $SU(4)_{F}$ flavor symmetry ineffective, and the failure of the heavy quark expansion due to the insufficiently large $m_{c}$. The increasing complexity of these decays further causes the ineffectiveness of factorization methods \cite{Bjorken:1988ya}. To overcome these challenges, alternative approaches for charmed hadron decays have been explored in various studies~\cite{Cheng:1991sn,Cheng:1995fe,Cheng:1993gf,Zenczykowski:1993hw,Fayyazuddin:1996iy,Dhir:2015tja,Cheng:2018hwl}. These approaches recognize the necessity of considering non-factorizable effects. On the other hand, the $SU(3)_{F}$ flavor symmetry method has been tested as a useful tool both in the beauty and charmed hadron decays. Its feasibility has been established in charmed baryon two-body and three-body semileptonic charmed baryon weak decays~\cite{Savage:1989qr,Savage:1991wu,Pirtskhalava:2011va,Grossman:2012ry,Lu:2016ogy,Geng:2017esc,Geng:2017mxn,Wang:2017gxe,Geng:2018plk,Geng:2018bow,Geng:2018rse,Geng:2018upx,Hsiao:2019yur,Geng:2019bfz,Jia:2019zxi,Geng:2019xbo,Liu:2023dvg}.

To investigate the non-resonant weak decays of $\mathbf{B}_{c}\rightarrow\mathbf{B}_{n}PP^{\prime}$, we make the assumption that the final state configurations of the pseudoscalar meson-pairs are predominantly characterized by S-wave ones. We express the decay amplitudes in terms of parity-conserving and violating components under $SU(3)_{F}$, following a similar framework as outlined in Ref.~\cite{Geng:2019xbo}. Building upon a thorough discussion~\cite{Geng:2018upx,Cen:2019ims} of the contribution from the color-antisymmetric part of the effective Hamiltonian
associated with the irreducible representation $\bar{\mathbf{6}}$ under $SU(3)_f$
 and incorporating additional experimental data, we extend our analysis to consider complete effective Hamiltonian contributions related to both 
 $\bar{\mathbf{6}}$ and $\mathbf{15}$ representations, resulting in 16 real parameters to be fitted with 28 available experiment data points. Furthermore, we discuss the possible error sources and some interesting findings of the new fit.

Our paper is organized as follows. We interpret the formalism and give the explicit amplitudes of all decay channels of $\mathbf{B}_{c}\rightarrow\mathbf{B}_{n}PP^{\prime}$ under the $SU(3)_{F}$ flavor symmetry in Sec.~\ref{Formalism}. In Sec.~\ref{Numerical results}, we present our numerical fitting results and discussions. Our conclusion is given in Sec.~\ref{Discussions and conclusions}.

\section{Formalism}\label{Formalism}
The nonleptonic three-body charmed baryon weak decays of $\mathbf{B}_{c}\rightarrow\mathbf{B}_{n}PP^{\prime}$ can be proceeded through the charmed quark decays of $c\rightarrow su\bar{d}, c\rightarrow ud\bar{d}\,(us\bar{s})$ and $c\rightarrow du\bar{s}$. Accordingly, the effective Hamiltonian at tree level is given by \cite{Buras:1998raa}
\begin{equation}\label{eq1}
    \mathcal{H}_{eff}=\frac{G_{F}}{\sqrt{2}}\sum_{i=-,+}[V_{c s} V^{*}_{u d}c_{i} O^{ds}_{i}+V_{c d} V^{*}_{u d}c_{i}\,(O^{dd}_{i}-O^{ss}_{i})+V_{c d} V^{*}_{u s}c_{i} O^{sd}_{i}]
\end{equation}
with the four-quark operators written as: 
\begin{equation}\label{eq2}
    O^{q_{1}q_{2}}_{\pm}=\frac{1}{2}\left[(\bar{u}q_{1})_{V-A}(\bar{q_{2}}c)_{V-A} \pm(\bar{q_{2}}q_{1})_{V-A}(\bar{u}c)_{V-A}\right]
\end{equation}
where $G_{F}$ is Fermi constant and $c_{i}$ represent the Wilson coefficients. The four quark operators $O^{ds}_{\pm},\,O^{dd}_{\pm}-O^{ss}_{\pm},\,O^{sd}_{\pm}$ are classified into so-called Cabibbo-favored$\,$(CF), singly Cabibbo-suppressed$\,$(CS) and doubly Cabibbo-suppressed$\,$(DCS) processes, respectively.

The three modes of charmed quark decays can be written as $c\rightarrow q^{i}q^{j}\,\overline{q}_{k}$ with $q_{i}=(u,d,s)$ is the triplet of light quarks under the $SU(3)_{F}$ flavor symmetry. The form of $q^{i}q^{j}\,\overline{q}_{k}$ can be decomposed as the irreducible representations of $\mathbf{3}\otimes\mathbf{3}\otimes\mathbf{\bar{3}}=\mathbf{15}\oplus\mathbf{\bar{6}}\oplus\mathbf{3}\oplus\mathbf{3}$, in which $\mathbf{15}$ and $\mathbf{\bar{6}}$ correspond to the color-symmetric operator $O^{q_{1}q_{2}}_{+}$ and the color-antisymmetric operator $O^{q_{1}q_{2}}_{-}$~\cite{Savage:1989qr,Savage:1991wu}, respectively. Consequently, the effective Hamiltonian can be divided into the symmetric part $H(\mathbf{15})$ and antisymmetric part $H(\bar{\mathbf{6}})$, defined as
\begin{equation}
    \mathcal{H}_{eff}= \frac{G_{F}}{\sqrt{2}}\left(c_{+} H(\mathbf{1 5})_{k}^{i j}+c_{-} H(\overline{\mathbf{6}})_{l k}\, \epsilon^{l i j}\right)
\left(\bar{q}_{i} q^{k}\right)_{V-A}\left(\bar{q}_{j} c\right)_{V-A}.
\end{equation}
Under $SU(3)_{F}$, the three lowest-lying charmed baryon states of $\mathbf{B}_{c}$ form anti-triplet charmed baryon states, and $\mathbf{B}_{n}$ and $P$ belong to octet baryon and pseudoscalar meson states. In this work, we adopt the same convention for the $SU(3)_{F}$ tensors 
as those in Refs.~\cite{Geng:2018upx,Cen:2019ims}. 

We assume that S-wave$\,(L=0)$ pseudoscalar meson pairs dominate in the non-resonant amplitudes. 
The decay amplitude can be written as
\begin{equation}\label{7}
    \mathcal{M}\left(\mathbf{B}_{c} \rightarrow \mathbf{B}_{n} P P^{\prime}\right) =\left\langle P^{\prime}P\mathbf{B}_{n}  \left|\mathcal{H}_{e f f}\right| \mathbf{B}_{c}\right\rangle=i \bar{u}_{\mathbf{B}_{n}}\left(A-B \gamma_{5}\right) u_{\mathbf{B}_{c}},
\end{equation}
where $u_{\mathbf{B}_{c,n}}$ are Dirac spinors of baryons,  and $A$ and $B$ represent the parity conserving and parity violating parts, respectively.
Assuming the dominance of $H(\overline{{\bf 6}})$ over $H({\bf 15})$, 
 $\langle\mathbf{B}_{n} P P^{\prime}\left|\mathcal{H}_{e f f}\right| \mathbf{B}_{c}\rangle$ contains six $SU(3)_F$ parameters in $A$ and $B$ amplitudes to be fitted with data~\cite{Cen:2019ims}. Due to the limitation of data points, we assume that final-state interactions (FSIs) are negligible between non-resonance states, and the parameters $A$ and $B$ are considered to be relatively real. 

\begin{figure}
\begin{minipage}{0.3\linewidth}
\begin{adjustbox}{width=\linewidth}
\begin{tikzpicture}
\begin{feynman}  
    \vertex (a1) {\(i\)};
    \vertex[right=2cm of a1] (a2);
    \vertex[right=2cm of a2] (a3) {\(i\)};
    
    \vertex[below=2em of a1] (b1) {\(j\)};
    \vertex[right=2cm of b1] (b2);
    \vertex[right=2cm of b2] (b3) {\(j\)};

    \vertex[below=2em of b1] (c1) {\(q_{c}\)};
    \vertex[right=2cm of c1] (c2);
    \vertex[right=2cm of c2] (c3) {\(l\)};
    
    \vertex[below=2em of c3] (d1) {\(m\)};
    \vertex[below=4em of d1] (d3) {\(\overline{k}\)};
    \vertex at ($(d1)!0.5!(d3) - (1.5cm, 0)$) (d2);
    
    \vertex[below=1em of d1] (e1) {\(\overline{n}\)};
    \vertex[below=2em of e1] (e3) {\(n\)};
    \vertex at ($(e1)!0.5!(e3) - (1cm, 0)$) (e2);
    
    \diagram* {
      {[edges=fermion]
        (a1) -- (a2) -- (a3),
        (b1) -- (b2) -- (b3),
        (c1) -- (c2) -- (c3),
      },
      (c2) -- [boson,bend right, edge label=\(W^+\)] (d2),
      (d3) -- [fermion, out=180, in=-45] (d2) -- [fermion, out=45, in=180] (d1),
      (e1) -- [fermion, out=-180, in=45] (e2) -- [fermion, out=-45, in=180] (e3)
    };

    \draw [decoration={brace}, decorate] ($(c1.south west)+(0.2em,0em)$) -- ($(a1.north west)+(0em,0em)$)
          node [pos=0.5, left] {\(\mathbf{B}_{c}\)};
    \draw [decoration={brace}, decorate] ($(d1.north east)+(0.0em,0em)$) -- ($(e1.south east)+(0.12em,0em)$)
          node [pos=0.5, right] {\(P\)};
    \draw [decoration={brace}, decorate] ($(e3.north east)+(0em,0em)$) -- ($(d3.south east)+(0em,0em)$)
         node [pos=0.5, right] {\(P^{\prime}\)};
    \draw [decoration={brace}, decorate] ($(a3.north east)+(0.05em,0em)$) -- ($(c3.south east)+(0em,0em)$)node [pos=0.5, right] {$\mathbf{B}_{n}$};
  \end{feynman}
  \end{tikzpicture}
\end{adjustbox}
\put(-80,-0){($a$)}
\end{minipage}
\begin{minipage}{0.3\linewidth}
\begin{adjustbox}{width=\linewidth}
\begin{tikzpicture}
\begin{feynman}  
    \vertex (a1) {\(q_{c}\)};
    \vertex[right=2cm of a1] (a2);
    \vertex[right=2cm of a2] (a3) {\(m\)};

    \vertex[below=1em of a3] (b1) {$\overline{k}$};
    \vertex[below=2em of b1] (b3) {\(l\)};
    \vertex at ($(b1)!0.5!(b3) - (1cm, 0)$) (b2);
    
    \vertex[below=1em of b3] (c1) {$\overline{n}$};
    \vertex[below=2em of c1] (c3) {\(n\)};
    \vertex at ($(c1)!0.5!(c3) - (1cm, 0)$) (c2);
    
    \vertex[below=7em of a1] (d1) {\(i\)};
    \vertex[right=2cm of d1] (d2);
    \vertex[right=2cm of d2] (d3) {\(i\)};

    \vertex[below=2em of d1] (e1) {\(j\)};
    \vertex[right=2cm of e1] (e2);
    \vertex[right=2cm of e2] (e3) {\(j\)};
    
    \diagram* {
      {[edges=fermion]
        (a1) -- (a2) -- (a3),
        (d1) -- (d2) -- (d3),
        (e1) -- (e2) -- (e3),
      },
      (a2) -- [boson,bend right, edge label=\(W^+\)] (b2),
      (c1) -- [fermion, out=-180, in=45] (c2) -- [fermion, out=-45, in=180] (c3),
      (b1) -- [fermion, out=-180, in=45] (b2) -- [fermion, out=-45, in=180] (b3)
    };

    \draw [decoration={brace}, decorate] (e1.south west) -- ($(a1.north west)+(0.2em,0em)$)
          node [pos=0.5, left] {\(\mathbf{B}_{c}\)};
    \draw [decoration={brace}, decorate] ($(a3.north east)+(0em,0em)$) -- ($(b1.south east)+(0.15em,0em)$)
          node [pos=0.5, right] {\(P\)};
    \draw [decoration={brace}, decorate] ($(b3.north east)+(0.2em,0em)$) -- ($(c1.south east)+(0em,0em)$)
         node [pos=0.5, right] {\(P^{\prime}\)};
    \draw [decoration={brace}, decorate] ($(c3.north east)+(0em,0em)$) -- ($(e3.south east)+(0.2em,0em)$)node [pos=0.5, right] {$\mathbf{B}_{n}$};
  \end{feynman}
  \end{tikzpicture}
\end{adjustbox}
\put(-80,-13){($b$)}
\end{minipage} \\ 
\begin{minipage}[b]{0.3\linewidth}
\begin{adjustbox}{width=\linewidth}
\begin{tikzpicture}
\begin{feynman}  
    \vertex (a1) {\(i\)};
    \vertex[right=2cm of a1] (a2);
    \vertex[right=2cm of a2] (a3) {\(i\)};

   \vertex[below=2em of a1] (b1) {\(q_{c}\)};
    \vertex[right=2cm of b1] (b2);
    \vertex[right=2cm of b2] (b3) {\(l\)};
    
    \vertex[below=1em of b3] (c1) {$n$};
    \vertex[below=2em of c1] (c3) {$\overline{n}$};
    \vertex at ($(c1)!0.5!(c3) - (1cm, 0)$) (c2);
    
    \vertex[below=1em of c3] (d1) {$m$};
    \vertex[below=2em of d1] (d3) {$\overline{k}$};
    \vertex at ($(d1)!0.5!(d3) - (1cm, 0)$) (d2);

    \vertex[below=7em of b1] (e1) {\(j\)};
    \vertex[right=2cm of e1] (e2);
    \vertex[right=2cm of e2] (e3) {\(j\)};
    
    \diagram* {
      {[edges=fermion]
        (a1) -- (a2) -- (a3),
        (b1) -- (b2) -- (b3),
        (e1) -- (e2) -- (e3),
      },
      (b2) -- [boson,bend right, edge label=\(W^+\)] (d2),
      (d3) -- [fermion, out=180, in=-45] (d2) -- [fermion, out=45, in=-180] (d1),
      (c3) -- [fermion, out=180, in=-45] (c2) -- [fermion, out=45, in=-180] (c1)
    };

    \draw [decoration={brace}, decorate] (e1.south west) -- ($(a1.north west)+(-0.1em,0em)$)
          node [pos=0.5, left] {\(\mathbf{B}_{c}\)};
    \draw [decoration={brace}, decorate] ($(c3.north east)+(0.16em,0em)$) -- ($(d1.south east)+(0.0em,0em)$)
          node [pos=0.5, right] {\(P_{1}\)};
    \draw [decoration={brace}, decorate] ($(d3.north east)+(0.07em,0em)$) -- ($(e3.south east)+(0em,0em)$)
         node [pos=0.5, right] {\(P_{1}^{\prime}\)};
    \draw [decoration={brace}, decorate] ($(a3.north east)+(0.15em,0em)$) -- ($(c1.south east)+(0em,0em)$)node [pos=0.5, right] {$\mathbf{B}_{n}$};
  \end{feynman}
  \end{tikzpicture}
\end{adjustbox}
\put(-80,-15){($c_{1}$)}
\end{minipage}
\begin{minipage}[b]{0.3\linewidth}
\begin{adjustbox}{width=\linewidth}
\begin{tikzpicture}
\begin{feynman}
    \vertex (a1) {\(q_{c}\)};
    \vertex[right=2cm of a1] (a2);
    \vertex[right=2cm of a2] (a3){\(l\)};
    
    \vertex[above=4em of a3] (b1) {\(m\)};
    \vertex[below=2em of b1] (b3) {$\overline{k}$};
    \vertex at ($(b1)!0.5!(b3) - (1cm, 0)$) (b2);
    
    \vertex[below=2em of a1] (c1) {\(i\)};
    \vertex[right=2cm of c1] (c2) ;
    \vertex[right=2cm of c2] (c3){\(i\)};

    \vertex[below=1em of c3] (d1) {\(n\)};
    \vertex[below=2em of d1] (d3) {$\overline{n}$};
    \vertex at ($(d1)!0.5!(d3) - (1cm, 0)$) (d2);

    \vertex[below=4em of c1] (e1) {\(j\)};
    \vertex[right=2cm of e1] (e2);
    \vertex[right=2cm of e2] (e3) {\(j\)};
    
    \diagram* {
      {[edges=fermion]
        (a1) -- (a2) -- (a3),
        (c1) -- (c2) -- (c3),
        (e1) -- (e2) -- (e3),
      },
      (b2) -- [boson,bend right, edge label=\(W^+\)] (a2),
      (b3) -- [fermion, out=180, in=-45] (b2) -- [fermion, out=45, in=180] (b1),
      (d3) -- [fermion, out=180, in=-45] (d2) -- [fermion, out=45, in=180] (d1)
    };

    \draw [decoration={brace}, decorate] ($(e1.south west)+(0.0em,0em)$) -- ($(a1.north west)+(0.2em,0em)$)
          node [pos=0.5, left] {\(\mathbf{B}_{c}\)};
    \draw [decoration={brace}, decorate] (a3.north east)+(0.18em,0em) -- (d1.south east)
          node [pos=0.5, right] {\(\mathbf{B}_{n}\)};
    \draw [decoration={brace}, decorate] (d3.north east)+(0.05em,0em) -- ($(e3.south east)+(0.0em,0em)$)
         node [pos=0.5, right] {\(P_{2}\)};
    \draw [decoration={brace}, decorate] (b1.north east)+(0.0em,0em) -- ($(b3.south east)+(0.12em,0em)$)
         node [pos=0.5, right] {\(P_{2}^{\prime}\)};
  \end{feynman}
  \end{tikzpicture}
\end{adjustbox}
\put(-80,-15){($c_{2}$)}
\end{minipage}
\begin{minipage}[b]{0.3\linewidth}
\begin{adjustbox}{width=\linewidth}
\begin{tikzpicture}
  \begin{feynman}
    \vertex (a1) {\(q_{c}\)};
    \vertex[right=2cm of a1] (a2);
    \vertex[right=2cm of a2] (a3){$m$};

    \vertex[below=1em of a3] (b1) {$\overline{l}$};
    \vertex[below=2em of b1] (b3) {$l$};
    \vertex at ($(b1)!0.5!(b3) - (1cm, 0)$) (b2);

    \vertex[below=1em of b3] (c1) {$\overline{o}$};
    \vertex[below=2em of c1] (c3) {$o$};
    \vertex at ($(c1)!0.5!(c3) - (1cm, 0)$) (c2);
    
    \vertex[below=7em of a1] (d1) {$i$};
    \vertex[right=2cm of d1] (d2);
    \vertex[right=2cm of d2] (d3) {$n$};
    
    \vertex[below=2em of d1] (e1) {$j$};
    \vertex[right=2cm of e1] (e2);
    \vertex[right=2cm of e2] (e3) {$j$};
    
    \diagram* {
      {[edges=fermion]
        (a1) -- (a2) -- (a3),
        (d1) -- (d2) -- (d3),
        (e1) -- (e2) -- (e3),
      },

      (a2) -- [boson, edge label=$W^{+}$] (d2),
      (b1) -- [fermion, out=-180, in=45] (b2) -- [fermion, out=-45, in=180] (b3),
      (c1) -- [fermion, out=-180, in=45] (c2) -- [fermion, out=-45, in=180] (c3)
    };

    \draw [decoration={brace}, decorate] (e1.south west) -- ($(a1.north west)+(0.25em,0em)$)
          node [pos=0.5, left] {$\mathbf{B}_{c}$};
    \draw [decoration={brace}, decorate] (a3.north east) -- ($(b1.south east)+(0.2em,0em)$)
          node [pos=0.5, right] {$P$};
    \draw [decoration={brace}, decorate] ($(b3.north east)+(0.15em,0em)$) -- (c1.south east)
          node [pos=0.5, right] {$P^{\prime}$};
    \draw [decoration={brace}, decorate] ($(c3.north east)+(0.0em,0em)$) -- ($(e3.south east)+(0.1em,0em)$)
          node [pos=0.5, right] {$\textbf{B}_{n}$};
  \end{feynman}
  \end{tikzpicture}
    \end{adjustbox}
   \put(-80,-15){($e$)}
\end{minipage}\hfill
\caption{Topology diagrams contributed by $H(\mathbf{15})$ before integrating the W boson, where ($a$), ($b$), ($c_{1}$) and ($c_{2}$) are W-emission processes, while ($e$) is an example of W-exchange processes. The parameters $a_{7}$, $a_{8}$, $a_{9}$, $a_{10}$ are given by $(a)$, $(b)$, $(c_{1})$ and $(c_{2})$, respectively.}
\label{fig1}
\end{figure}
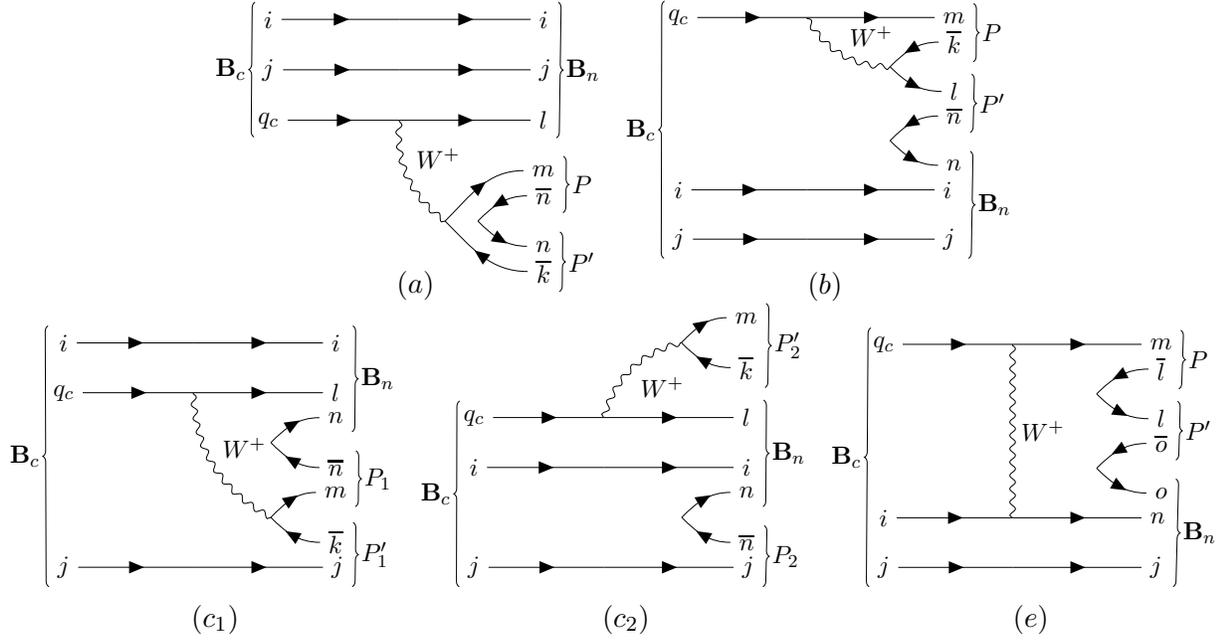

However, the $H(\mathbf{\bar{6}})$ fitting~\cite{Cen:2019ims} has presented large deviations with updated experiment data, such as $\Lambda^{+}_{c}\rightarrow\Lambda^{0}K^{+}\pi^{0}$~\cite{BESIII:2023sdr}, and $\Xi^{0}_{c}\rightarrow\Xi^{0}K^{+}K^{-}$~\cite{Belle:2020ito}, indicating the $H(\mathbf{15})$ contribution is of great significance in the $SU(3)_{F}$ fitting. In this study, we first incorporate $H(\mathbf{15})$ into the analysis of charmed baryon three-body weak decay. We list all possible topological diagrams contributed by $H(\mathbf{15})$ before integrating the $W$ boson in Fig.~\ref{fig1}. 
Focusing on \(H(\mathbf{15})\), the quarks represented by \(q_c\) and \(i\) of the W-exchange part in Fig.~1$e$ are symmetric in color. Meanwhile, \(q_c\) and \(i\) originate from \(\mathbf{B}_c\), where the color of the quarks is totally antisymmetric. We conclude that the topology of W-exchange processes like Fig.~1$e$ does not contribute to \(H(\mathbf{15})\), owing to the K\"orner-Pati-Woo theorem~\cite{Korner:1970xq,Pati:1970fg}.
From these diagrams we can get complete $A$ and $B$ $SU(3)_{F}$ amplitudes, given by
\begin{equation}\label{eq9}
\begin{aligned}
A\left(\mathbf{B}_{c} \rightarrow \mathbf{B}_{n} P P^{\prime}\right)
= a_{1}\left(\overline{\mathbf{B}}_{n}\right)_{\;\,i\;\,}^{\;\,k\;\,}(\overline{P})_{l}^{m}(\overline{P})_{m}^{l} H(\mathbf{\bar{6}})_{\,j\,k\,} T^{ij}+a_{2}\left(\overline{\mathbf{B}}_{n}\right)_{\;\,i\;\,}^{\;\,k\;\,}(\overline{P})_{j}^{m}(\overline{P})_{m}^{l} H(\mathbf{\bar{6}})_{\,k\;l}\; &T^{ij} \\
+a_{3}\left(\overline{\mathbf{B}}_{n}\right)_{\;\,i\;\,}^{\;\,k\;\,}(\overline{P})_{k}^{m}(\overline{P})_{m}^{l} H(\mathbf{\bar{6}})_{\,j\;l\,} T^{ij} 
+a_{4}\left(\overline{\mathbf{B}}_{n}\right)_{\;\,i\;\,}^{\;\,k\;\,}(\overline{P})_{\;j\;}^{\;l\;}(\overline{P})_{k}^{m} H(\mathbf{\bar{6}})_{\,l\,m} &T^{i j} \\
+a_{5}\left(\overline{\mathbf{B}}_{n}\right)_{\;\,k\;\,}^{\;\,l\;\,}(\overline{P})_{j}^{m}(\overline{P})_{m}^{k} H(\mathbf{\bar{6}})_{\;i\;l\;} T^{i j} 
+a_{6}\left(\overline{\mathbf{B}}_{n}\right)_{\;\,k\;\,}^{\;\,l\;\,}(\overline{P})_{j}^{m}(\overline{P})_{\,l\;}^{\,k\;} H(\mathbf{\bar{6}})_{\,i\,m} &T^{i j} \\
+a_{7}\left(\overline{\mathbf{B}}_{n}\right)_{i j l}(\overline{P})_{m}^{n}(\overline{P})_{n}^{k} H(\mathbf{15})_{k}^{l m} T^{i j}
+a_{\;8\,}\left(\overline{\mathbf{B}}_{n}\right)_{i j n}(\overline{P})_{m}^{k}(\overline{P})_{l}^{n} H(\mathbf{15})_{k}^{l m} &T^{i j}\\
+a_{9}\left(\overline{\mathbf{B}}_{n}\right)_{i l n}(\overline{P})_{j}^{k}(\overline{P})_{m}^{n}H(\mathbf{15})_{k}^{l m} T^{i j}
+a_{10}\left(\overline{\mathbf{B}}_{n}\right)_{i l n}(\overline{P})_{m}^{k}(\overline{P})_{j}^{n}H(\mathbf{15})_{k}^{l m} &T^{i j}\\
B\left(\mathbf{B}_{\mathbf{c}} \rightarrow \mathbf{B}_{n} P P^{\prime}\right)
=A\left(\mathbf{B}_{\mathbf{c}} \rightarrow \mathbf{B}_{n} P P^{\prime}\right)\left\{a_{i} \rightarrow b_{i}\right\}\qquad\qquad\qquad\qquad\qquad\qquad\quad\qquad\;&  
\end{aligned}
\end{equation}
where $(\overline{\mathbf{B}}_{n})_{i j l}=\epsilon_{i j k}\,(\overline{\mathbf{B}}_{n})^{k}_{l}$. 
We observe that in Figs. 1$a$ and 1$b$, \(\mathbf{B}_n\) is composed of the two spectator quarks from \(\mathbf{B}_c\), whereas in Figs.~1$c_1$ and 1$c_2$, it consists of only one spectator quark. Given that the light quarks in \(\mathbf{B}_c\) and \(\mathbf{B}_n\) have similar wave functions in the quark model, it is reasonable to assume that the contributions from ($c_1$) and ($c_2$) are smaller than those from ($a$) and ($b$). Hence,  we take 
$a_{9,10}$ and $b_{9,10}$, which come from the topology of Figs.~1$c_1$ and 1$c_2$,
 to be zero in the following. 

The explicit full expansions of $A\,(\Lambda^{+}_{c}\rightarrow \mathbf{B}_{n}PP^{\prime})$, $A\,(\Xi^{+}_{c}\rightarrow \mathbf{B}_{n}PP^{\prime})$, $A\,(\Xi^{0}_{c}\rightarrow \mathbf{B}_{n}PP^{\prime})$  within 16 parameters are presented in Table \ref{tab:A-amplitudes1} to Table \ref{tab:A-amplitudes3.2} in Appendix \ref{appendix}, while $B\,(\mathbf{B}_{c}\rightarrow \mathbf{B}_{n}PP^{\prime})$ can be simply acquired by the replacement of $A\,(\mathbf{B}_{c}\rightarrow \mathbf{B}_{n}PP^{\prime})\,\{a_{i}\rightarrow b_{i},\;i=1\sim8\}$. As we only consider the physical quantities after integrating over the phase space, we assume the amplitudes of $a_i$ and $b_i$ to be independent of $m_{23}^2$, which can be justified in the limit of the $SU(3)_f$ flavor symmetry
\footnote{We have examined the simplest first-order of $m_{23}^{2}$ dependency in the form of $a_{i}=a^{0}_{i}\,[\,1+(m^{2}_{23}/m^{2}_{\mathbf{B}_{c}})\,r\,]$. The fitting result closely mirrors that presented in this paper  with $r=0.09$. }. By introducing the kinematic correction $\kappa\,(m^{2}_{23})$, the differential decay width and averaged up-down asymmetry of $\mathbf{B}_{c}\rightarrow\mathbf{B}_{n}PP^{\prime}$ can be derived as ~\cite{Cen:2019ims}
\begin{equation}\label{eq6}
    \Gamma=\int_{m_{12}^{2}} \int_{m_{23}^{2}} \frac{1}{(2 \pi)^{3}}\frac{A^{2}+\kappa^{2}\left(m^{2}_{23}\right) B^{2}}{32 m_{\mathbf{B}_{\mathbf{c}}}^{3}} d m_{12}^{2} d m_{23}^{2}\,,
\end{equation}
and
\begin{equation}
    \langle\alpha\rangle \equiv \frac{2\int_{m_{12}^{2}} \int_{m_{23}^{2}} \kappa\left(m^{2}_{23}\right)AB\,\, d m_{12}^{2} d m_{23}^{2}}{\int_{m_{12}^{2}} \int_{m_{23}^{2}} A^{2}+\kappa^{2}\left(m^{2}_{23}\right)B^{2}\, d m_{12}^{2} d m_{23}^{2}}\,,
\end{equation}
respectively. The kinematic correction $\kappa\,(m^{2}_{23})$ is defined as
\begin{equation}
    \kappa\,\left(m^{2}_{23}\right)=\sqrt{\frac{\left(m_{\mathbf{B}_{c}}-m_{\mathbf{B}_{n}}\right)^{2}-m^{2}_{23}}{\left(m_{\mathbf{B}_{c}}+m_{\mathbf{B}_{n}}\right)^{2}-m^{2}_{23}}}
\end{equation}
with $m_{23}$ is the sum of the 3-momentum of the two pseudoscalar mesons in the rest frame.
\section{Numerical results}\label{Numerical results}
We make use of the minimum $\chi^{2}$ fit in the numerical analysis to obtain the values of 16 parameters $a_{i}$ and $b_{i}$ in Eq.~(\ref{eq9}) under $SU(3)_{F}$ for $\mathbf{B}_{c}\rightarrow\mathbf{B}_{n}PP^{\prime}$. The validity can be tested via $\chi^{2}/d.o.f$. The minimum $\chi^{2}$ fit approach is given by
\begin{equation}
    \chi^{2}=\sum_{i}\left(\frac{\mathcal{B}_{S U(3)}^{\,i}-\mathcal{B}_{\text {data }}^{\,i}}{\sigma_{\text {data }}^{\,i}}\right)^{2}
\end{equation}
in which $\mathcal{B}_{S U(3)}^{\,i}$ is the $i-$th decay branching ratio from $SU(3)_{F}$ fitting predictions, $\mathcal{B}_{\text {data }}^{\,i}$ represents the $i-$th experiment data, and $\sigma_{\text {data }}^{\,i}$ stands for the $i-$th experiment error, while $i=1,2,...,28$ for 28 experiment measured channels in Table \ref{tab:experiment and reproductions}.  

\begin{table}[H]
    \centering
    \caption{The experimental data from Refs.~\cite{Workman:2022ynf,BESIII:2018qyg,BESIII:2023rky,LHCb:2017xtf,BESIII:2020kzc,Belle:2019bgi,Belle:2018kzz,Belle:2013ntc,Belle:2013jfq,Belle:2020xku,BESIII:2023dvx,Belle:2022pwd,BESIII:2023pia,BESIII:2023sdr,BESIII:2015bjk,CLEO:1995rjm,LHCb:2020gge,SELEX:2008rjp,FOCUS:2003gpe,Belle:2020ito} and reproductions for $\mathcal{B}(\mathbf{B}_{c}\rightarrow\mathbf{B}_{n}PP^{\prime})$}
    
\renewcommand{\arraystretch}{1}
\footnotesize
\begin{tabular}{lcclcc}
\hline
$\qquad\qquad$Channels$\qquad$&Data&Our fittings&$\qquad\qquad$Channels$\qquad$&Data&Our fittings\\
\hline
$10^{2}\mathcal{B}(\Lambda^{+}_{c}\rightarrow p\pi^{+}K^{-})$&$3.4\pm0.4$&$3.4\pm0.4$&$10^{2}\mathcal{B}(\Lambda^{+}_{c}\rightarrow \Sigma^{+}\pi^{0}\pi^{0})$&$1.3\pm0.1$&$1.3\pm0.1$\\
$10^{3}\mathcal{B}(\Lambda^{+}_{c}\rightarrow \Lambda^{0}K^{+}\bar{K}^{0})$&$5.6\pm1.1$&$5.9\pm1.0$&$10^{4}\mathcal{B}(\Lambda^{+}_{c}\rightarrow p\pi^{-}K^{+})$&$1.0\pm0.1$&$1.0\pm0.1$
\\
$10^{2}\mathcal{B}(\Lambda^{+}_{c}\rightarrow \Lambda^{0}\pi^{+}\eta^{0})$&$1.8\pm0.3$&$1.9\pm0.3$&$10^{4}\mathcal{B}(\Lambda^{+}_{c}\rightarrow nK^{+}\bar{K}^{0})$&$8.6^{+3.8}_{-3.0}$&$6.5\pm2.2$\\
$10^{3}\mathcal{B}(\Lambda^{+}_{c}\rightarrow \Lambda^{0}\pi^{0}K^{+})$&$1.5\pm0.3$&$1.4\pm0.3$&$10^{2}\mathcal{B}(\Lambda^{+}_{c}\rightarrow p\pi^{0}K^{0}_{S})$&$1.9\pm0.1$&$1.9\pm0.1$\\
$10^{2}\mathcal{B}(\Lambda^{+}_{c}\rightarrow \Sigma^{+}\pi^{+}\pi^{-})$&$2.9\pm0.5$&$2.8\pm0.5$&$10^{2}\mathcal{B}(\Lambda^{+}_{c}\rightarrow n\pi^{+}K^{0}_{S})$&$1.9\pm0.1$&$1.9\pm0.1$\\
$10^{2}\mathcal{B}(\Lambda^{+}_{c}\rightarrow \Sigma^{-}\pi^{+}\pi^{+})$&$1.9\pm0.2$&$2.0\pm0.2$&$10^{2}\mathcal{B}(\Xi^{+}_{c}\rightarrow \Sigma^{+}\pi^{+}K^{-})$&$2.6\pm1.2$&$3.9\pm0.4$\\
$10^{2}\mathcal{B}(\Lambda^{+}_{c}\rightarrow \Sigma^{0}\pi^{+}\pi^{0})$&$2.2\pm0.8$&$1.0\pm0.1$&$10^{2}\mathcal{B}(\Xi^{+}_{c}\rightarrow \Xi^{0}\pi^{+}\pi^{0})$&$6.7\pm3.5$&$1.0\pm0.3$\\
$10^{3}\mathcal{B}(\Lambda^{+}_{c}\rightarrow \Sigma^{0}\pi^{+}\eta^{0})$&$8.2\pm0.9$&$8.3\pm0.8$&$10^{3}\mathcal{B}(\Xi^{+}_{c}\rightarrow \Sigma^{+}\pi^{+}\pi^{-})$&$14.0\pm8.0$&$6.5\pm1.6$\\
$10^{3}\mathcal{B}(\Lambda^{+}_{c}\rightarrow \Sigma^{+}\pi^{-}K^{+})$&$2.0\pm0.4$&$1.6\pm0.3$&$10^{3}\mathcal{B}(\Xi^{+}_{c}\rightarrow \Sigma^{-}\pi^{+}\pi^{+})$&$5.1\pm3.4$&$6.9\pm2.3$\\
$10^{3}\mathcal{B}(\Lambda^{+}_{c}\rightarrow \Xi^{-}\pi^{+}K^{+})$&$3.3\pm0.9$&$1.5\pm0.5$&$10^{3}\mathcal{B}(\Xi^{+}_{c}\rightarrow \Sigma^{+}K^{+}K^{-})$&$4.2\pm2.5$&$0.4\pm0.2$\\
$10^{3}\mathcal{B}(\Lambda^{+}_{c}\rightarrow p\pi^{+}\pi^{-})$&$4.7\pm0.3$&$4.6\pm0.3$&$10^{2}\mathcal{B}(\Xi^{0}_{c}\rightarrow \Lambda^{0}\pi^{+}K^{-})$&$1.2\pm0.4$&$1.3\pm0.3$\\
$10^{4}\mathcal{B}(\Lambda^{+}_{c}\rightarrow pK^{+}K^{-})$&$5.2\pm1.2$&$4.8\pm1.0$&$10^{4}\mathcal{B}(\Xi^{0}_{c}\rightarrow \Lambda^{0}K^{+}K^{-})$&$5.1\pm1.9$&$4.5\pm0.7$\\
$10^{2}\mathcal{B}(\Lambda^{+}_{c}\rightarrow p\bar{K}^{0}\eta^{0})$&$0.8\pm0.2$&$0.7\pm0.1$&$10^{3}\mathcal{B}(\Xi^{0}_{c}\rightarrow \Xi^{0}K^{+}K^{-})$&$0.7\pm0.2$&$1.0\pm0.1$\\
$10^{3}\mathcal{B}(\Lambda^{+}_{c}\rightarrow \Xi^{0}\pi^{0}K^{+})$&$7.8\pm1.6$&$8.0\pm1.5$&$10^{2}\mathcal{B}(\Xi^{+}_{c}\rightarrow \Xi^{-}\pi^{+}\pi^{+})$&$2.9\pm1.3$&$4.0\pm1.1$\\
\hline
\end{tabular}

    \label{tab:experiment and reproductions}
\end{table}

\begin{table}[H]
    \centering
    \caption{Fitting values for $a_{i}$ and $b_{i}$ in unit of $\frac{G_{F}}{\sqrt{2}}\cdot$ GeV$^{2}$}
    
\renewcommand{\arraystretch}{1}
\footnotesize
\begin{tabular}{llllll}
\hline
$a_{i}$&$\qquad$Result&$\qquad b_{i}$&$\qquad$Result\\
\hline
$a_{1}$&$\qquad 3.12\pm0.99$&$\qquad b_{1}$&$\qquad 8.08\pm4.15$\\
$a_{2}$&$\qquad 0.06\pm0.61$&$\qquad b_{2}$&$\qquad 2.20\pm1.88$\\
$a_{3}$&$\qquad -4.91\pm1.09$&$\qquad b_{3}$&$\qquad -18.04\pm7.15$\\
$a_{4}$&$\qquad 2.05\pm0.44$&$\qquad b_{4}$&$\qquad -3.10\pm3.72$\\
$a_{5}$&$\qquad 6.36\pm0.76$&$\qquad b_{5}$&$\qquad -12.83\pm8.58$\\
$a_{6}$&$\qquad -1.09\pm0.89$&$\qquad b_{6}$&$\qquad 12.15\pm2.91$\\
$a_{7}$&$\qquad 1.69\pm1.40$&$\qquad b_{7}$&$\qquad -30.06\pm3.18$\\
$a_{8}$&$\qquad -5.22\pm0.37$&$\qquad b_{8}$&$\qquad -2.85\pm6.09$\\
\hline
\end{tabular}

    \label{tab:ai and bi}
\end{table}

We now discuss the data input in Table \ref{tab:experiment and reproductions}. Most importantly, the contributions from two-body resonances are excluded from these 28 data in the table. Aside from the usage of nonresonant experimental data, we also take $\Lambda^{+}_{c}\rightarrow\Sigma^{+}(\rho^{0}\rightarrow\pi^{+}\pi^{-})$ \cite{Geng:2020zgr,CLEO:1993cxp} and $\Lambda^{+}_{c}\rightarrow(\Xi(1530)^{+}\rightarrow\Xi^{-}\pi^{+})K^{+}$ \cite{BESIII:2018cvs} into consideration for subtracting the resonant contributions. To obtain absolute branching ratios, we incorporate specific branching ratio measurements, $\mathcal{B}\,(\Lambda^{+}_{c}\rightarrow p\pi^{+}K^{-})=(6.8\pm0.3)\%$ \cite{Belle:2013jfq} and $\mathcal{B}\,(\Xi^{0}_{c}\rightarrow \Xi^{-}\pi^{+})=(1.8\pm0.5)\%$ \cite{Belle:2018kzz}, provided by Belle. These measurements serve as crucial data  to convert relative branching ratios into absolute values.

The fitted 16 parameters are collected in Table \ref{tab:ai and bi} with $\chi^{2}/d.o.f=1.5$ and $d.o.f=12$ standing for the degree of freedom.  In Table \ref{tab:experiment and reproductions}, the branching ratios of 28 input data have been reproduced, and we can see that the $SU(3)_{F}$ fittings are in good agreement with the data. We list our all numerical fitting results for the branching ratios of $\Lambda^{+}_{c}\rightarrow\mathbf{B}_{n}PP^{\prime}$, $\Xi^{+}_{c}\rightarrow\mathbf{B}_{n}PP^{\prime}$ and $\Xi^{0}_{c}\rightarrow\mathbf{B}_{n}PP^{\prime}$ in Tables \ref{tab:Numerical result 1},~\ref{tab:Numerical result 2},~\ref{tab:Numerical result 3.1} and \ref{tab:Numerical result 3.2}, respectively.

Our predictions include the branching ratio of $\mathcal{B}\,(\Lambda^{+}_{c}\rightarrow\Sigma^{+}K^{+}K^{-})=(4.6\pm0.4)\times10^{-4}$, which is notably smaller than the  one of $(2.0\pm0.4)\times10^{-3}$ observed by BESIII \cite{BESIII:2023rky} but aligns with the upper limit prediction of $8\times10^{-4}$ from Belle \cite{Belle:2001hyr}. This suggests that resonant contributions still exist.
Working backwards, it suggests that 
 $\mathcal{B}\,(\Lambda^{+}_{c}\rightarrow(\Xi(1690)^{0}\rightarrow\Sigma^{+}K^{-})K^{+})$ is about $(1.5\pm0.4)\times10^{-3}$. Moreover, the predicted branching ratios for $\mathcal{B}\,(\Lambda^{+}_{c}\rightarrow\Sigma^{0}\pi^{+}\pi^{0})$ and $\mathcal{B}\,(\Lambda^{+}_{c}\rightarrow\Xi^{-}\pi^{+}K^{+})$ are approximately half the magnitude of their experimental measurements, hinting the presence of other resonant contributions from excited state particles. We emphasize the need for further experimental investigations to confirm their existences and extract the branch fractions of these states. 

Notably, our prediction of $\mathcal{B}\,(\Xi^{+}_{c}\rightarrow p\pi^{+}K^{-})=(28.3\pm2.6)\times10^{-3}$ significantly exceeds the results of $(11\pm4)\times10^{-3}$ from LHCb \cite{LHCb:2020gge} and $(4.5\pm2.2)\times10^{-3}$ from Belle \cite{Belle:2019bgi}. 
It is interesting to point out that 
$A\left(
\Xi^{+}_{c}\rightarrow p\pi^{+}K^{-}
\right) = A \left(
\Lambda^{+}_{c}\rightarrow\Sigma^{+}\pi^{-}K^{+}
\right)$ in the exact $SU(3)_{F}$ symmetry, but
their released energies are in great difference of 0.9~GeV and 0.5~GeV for the former and latter, respectively~\footnote{From Eq.~\eqref{eq6}, we see that $\Gamma\propto \Delta E ^4 $
if $A$ and $B$ are held as constant, 	
 with $\Delta E$ the released energy. }, leading to the  hierarchy of $\Gamma\,(\Xi_c^+ \to p \pi ^+ K^-) \approx 10\Gamma\, (\Lambda_c^+ \to \Sigma ^+ \pi ^- K^+)$. However, this analysis is  in sharp contrast to 
the experimental data, indicating that their decay widths are approximately the same. This implies a large $SU(3)_{F}$ flavor symmetry breaking effect, which may come from resonant hadrons. We strongly suggest that future experiments revise these two channels.
 
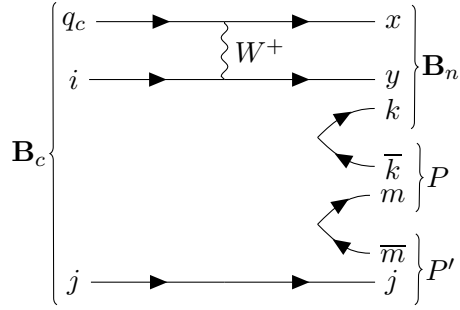
\begin{figure}
\begin{minipage}[t]{0.35\linewidth}
\begin{adjustbox}{width=\linewidth}
\begin{tikzpicture}
  \begin{feynman}
    \vertex (a1) {\(q_{c}\)};
    \vertex[right=2cm of a1] (a2);
    \vertex[right=2cm of a2] (a3){$x$};
    
    \vertex[below=2em of a1] (b1) {$i$};
    \vertex[right=2cm of b1] (b2);
    \vertex[right=2cm of b2] (b3) {$y$};

    \vertex[below=1em of b3] (c1) {$k$};
    \vertex[below=2em of c1] (c3) {$\overline{k}$};
    \vertex at ($(c1)!0.5!(c3) - (1cm, 0)$) (c2);

    \vertex[below=1em of c3] (d1) {$m$};
    \vertex[below=2em of d1] (d3) {$\overline{m}$};
    \vertex at ($(d1)!0.5!(d3) - (1cm, 0)$) (d2);
    
    \vertex[below=7em of b1] (e1) {$j$};
    \vertex[right=2cm of e1] (e2);
    \vertex[right=2cm of e2] (e3) {$j$};
    
    \diagram* {
      {[edges=fermion]
        (a1) -- (a2) -- (a3),
        (b1) -- (b2) -- (b3),
        (e1) -- (e2) -- (e3),
      },

      (a2) -- [boson, edge label=$W^{+}$] (b2),
      (c3) -- [fermion, out=180, in=-45] (c2) -- [fermion, out=45, in=180] (c1),
      (d3) -- [fermion, out=180, in=-45] (d2) -- [fermion, out=45, in=180] (d1)
    };
    \draw [decoration={brace}, decorate] (e1.south west) -- ($(a1.north west)+(0.15em,0em)$)
          node [pos=0.5, left] {$\mathbf{B}_{c}$};
    \draw [decoration={brace}, decorate] ($(c3.north east)+(0.2em,0em)$) -- ($(d1.south east)$)
          node [pos=0.5, right] {$P$};
    \draw [decoration={brace}, decorate] ($(d3.north east)+(0.0em,0em)$) -- ($(e3.south east)+(0.2em,0em)$)
          node [pos=0.5, right] {$P^{\prime}$};
    \draw [decoration={brace}, decorate] ($(a3.north east)+(0.0em,0em)$) -- ($(c1.south east)$)
          node [pos=0.5, right] {$\mathbf{B}_{n}$};
  \end{feynman}
  \end{tikzpicture}
    \end{adjustbox}
\end{minipage}
\caption{Topology diagram of parameter $a_{5}$}
\label{fig2}
\end{figure}

\begin{table}[H]
	\centering
	\caption{Numerical results for $\mathcal{B}(\Lambda^{+}_{c}\rightarrow\mathbf{B}_{n}PP^{\prime})$}
	
\renewcommand{\arraystretch}{1}
\footnotesize
\begin{tabular}{llllll}
\hline
CF mode$\qquad$ &
  $10^{2}\mathcal{B}$ &
  $\qquad$CF mode$\qquad$ &
  $10^{2}\mathcal{B}$
  \\ \hline
$\Lambda^{0}\pi^{+}\eta^{0}$ &
  $1.93\pm0.27$ &
  $\qquad$$p\pi^{0}\bar{K}^{0}$ &
   $3.90\pm0.29$ 
  \\
$\Lambda^{0}K^{+}\bar{K}^{0}$ &
  $0.59\pm0.10$ &
  $\qquad$$p\bar{K}^{0}\eta^{0}$ &
   $0.74\pm0.11$ 
  \\
$\Sigma^{0}\pi^{+}\pi^{0}$ &
  $1.01\pm0.10$ &
  $\qquad$$p\pi^{+}K^{-}$ &
  $3.42\pm0.36$ 
  \\
$\Sigma^{0}\pi^{+}\eta^{0}$ &
   $0.83\pm0.08$ &
   $\qquad$$n\pi^{+}\bar{K}^{0}$ &
  $3.67\pm0.22$ 
   \\
$\Sigma^{0}K^{+}\bar{K}^{0}$ &
  $(2.34\pm0.83)\times10^{-2}$ &
  $\qquad$$\Sigma^{+}\pi^{0}\pi^{0}$ &
   $1.33\pm0.10$ 
   \\
$\Sigma^{-}\pi^{+}\pi^{+}$ &
  $1.95\pm0.20$ &
  $\qquad$$\Sigma^{+}\pi^{0}\eta^{0}$ &
   $0.87\pm0.09$ 
  \\
$\Xi^{-}\pi^{+}K^{+}$ &
  $0.15\pm0.05$ &
  $\qquad$$\Sigma^{+}\eta^{0}\eta^{0}$ &
   $(4.29\pm4.04)\times10^{-6}$
   \\
$\Xi^{0}\pi^{0}K^{+}$ &
  $0.80\pm0.15$ &
  $\qquad$$\Sigma^{+}\pi^{+}\pi^{-}$ &
   $2.81\pm0.47$ 
   \\
$\Xi^{0}K^{+}\eta^{0}$ &
  $(1.45\pm0.32)\times10^{-2}$ &
  $\qquad$$\Sigma^{+}K^{+}K^{-}$ &
  $(4.58\pm0.38)\times10^{-2}$ 
  \\
$\Xi^{0}\pi^{+}K^{0}$ &
  $0.97\pm0.33$ &
  $\qquad$$\Sigma^{+}K^{0}\bar{K}^{0}$ &
  $(0.86^{+1.18}_{-0.86})\times10^{-2}$ 
  \\
\hline
\hline
CS mode$\qquad$ &
  $10^{4}\mathcal{B}$ &
  $\qquad$CS mode$\qquad$ &
  $10^{4}\mathcal{B}$ 
\\ 
\hline
  $\Lambda^{0}\pi^{0}K^{+}$ &
   $13.6\pm2.5$ &
  $\qquad$$p\eta^{0}\eta^{0}$ &
   $4.46\pm1.18$
  \\
  $\Lambda^{0}K^{+}\eta^{0}$ &
$0.59\pm0.24$ &
  $\qquad$$p\pi^{+}\pi^{-}$ &
   $46.4\pm3.0$
  \\
  $\Lambda^{0}\pi^{+}K^{0}$ &
   $39.3\pm7.6$ &
  $\qquad$$pK^{+}K^{-}$ &
   $4.82\pm1.05$
  \\
  $\Sigma^{0}\pi^{0}K^{+}$ &
   $8.24\pm1.41$ &
  $\qquad$$pK^{0}\bar{K}^{0}$ &
  $3.53\pm1.68$
   \\
  $\Sigma^{0}K^{+}\eta^{0}$ &
   $(4.29\pm1.23)\times10^{-2}$ &
  $\qquad$$n\pi^{+}\pi^{0}$ &
  $25.3\pm5.5$
   \\
  $\Sigma^{0}\pi^{+}K^{0}$ &
  $3.31\pm0.97$ &
  $\qquad$$n\pi^{+}\eta^{0}$ &
   $45.2\pm12.1$
  \\
  $\Sigma^{-}\pi^{+}K^{+}$ &
  $3.73\pm1.11$ &
  $\qquad$$nK^{+}\bar{K}^{0}$ &
   $6.54\pm2.17$
   \\
  $\Xi^{-}K^{+}K^{+}$ &
  $(3.33^{+4.23}_{-3.33})\times10^{-3}$ &
  $\qquad$$\Sigma^{+}\pi^{0}K^{0}$ &
  $3.46\pm1.01$
   \\
  $\Xi^{0}K^{+}K^{0}$ &
  $(1.24^{+1.62}_{-1.24})\times10^{-3}$ &
  $\qquad$$\Sigma^{+}K^{0}\eta^{0}$ &
  $(8.38\pm2.40)\times10^{-2}$
  \\
  $p\pi^{0}\pi^{0}$ &
   $39.6\pm9.2$ &
  $\qquad$$\Sigma^{+}\pi^{-}K^{+}$ &
  $16.1\pm3.0$
  \\
  $p\pi^{0}\eta^{0}$ &
   $25.8\pm6.1$ &
  $\qquad$ &
  $\quad$
  \\
\hline
\hline
  DCS mode$\qquad$ &
  $10^{6}\mathcal{B}$ &
  $\qquad$DCS mode$\qquad$ &
  $10^{6}\mathcal{B}$ 
  \\ 
  \hline
  $\Lambda^{0}K^{+}K^{0}$ &
  $7.22\pm1.08$ &
  $\qquad$$p\pi^{-}K^{+}$ &
  $96.1\pm9.2$
  \\
  $\Sigma^{0}K^{+}K^{0}$ &
  $1.10\pm0.42$ &
  $\qquad$$n\pi^{0}K^{+}$ &
   $79.1\pm28.8$
  \\
  $\Sigma^{-}K^{+}K^{+}$ &
  $1.08\pm0.41$ &
  $\qquad$$nK^{+}\eta^{0}$ &
   $0.85^{+1.08}_{-0.85}$
  \\
  $p\pi^{0}K^{0}$ &
   $56.8\pm15.8$ &
   $\qquad$$n\pi^{+}K^{0}$ &
  $153\pm61$ 
   \\
  $pK^{0}\eta^{0}$ &
   $13.6\pm2.6$ &
  $\qquad$$\Sigma^{+}K^{0}K^{0}$ &
  $1.09\pm0.42$ 
   \\
  \hline

\end{tabular}

	\label{tab:Numerical result 1}
\end{table}

\begin{table}[H]
	\centering
	\caption{Numerical results for $\mathcal{B}(\Xi^{+}_{c}\rightarrow\mathbf{B}_{n}PP^{\prime})$}
	
\renewcommand{\arraystretch}{1}
\footnotesize
\begin{tabular}{llllll}
\hline
CF mode$\qquad$ &
  $10^{2}\mathcal{B}$ &
  $\qquad$CF mode$\qquad$ &
  $10^{2}\mathcal{B}$
 \\ 
 \hline
$\Lambda^{0}\pi^{+}\bar{K}^{0}$ &
  $0.61\pm0.60$ &
  $\qquad$$\Xi^{0}K^{+}\bar{K}^{0}$ &
  $0.18\pm0.08$
  \\
$\Sigma^{0}\pi^{+}\bar{K}^{0}$ &
  $7.33\pm1.21$ &
 $\qquad$$p\bar{K}^{0}\bar{K}^{0}$ &
  $2.46\pm0.69$
  \\
$\Xi^{-}\pi^{+}\pi^{+}$ &
  $3.95\pm1.10$ &
  $\qquad$$\Sigma^{+}\pi^{0}\bar{K}^{0}$ &
  $5.96\pm1.08$
  \\
$\Xi^{0}\pi^{+}\pi^{0}$ &
  $1.01\pm0.28$ &
  $\qquad$$\Sigma^{+}\bar{K}^{0}\eta^{0}$ &
  $(3.02\pm1.78)\times10^{-2}$
  \\
$\Xi^{0}\pi^{+}\eta^{0}$ &
  $1.18\pm0.56$ &
  $\qquad$$\Sigma^{+}\pi^{+}K^{-}$ &
  $3.86\pm0.44$
  \\
  \hline
\hline
CS mode$\qquad$ &
  $10^{3}\mathcal{B}$ &
  $\qquad$CS mode$\qquad$ &
  $10^{3}\mathcal{B}$ 
  \\ \hline
$\Lambda^{0}\pi^{+}\pi^{0}$ &
  $0.85\pm0.17$ &
  $\qquad$$p\pi^{0}\bar{K}^{0}$ &
  $7.39\pm1.77$ 
  \\
$\Lambda^{0}\pi^{+}\eta^{0}$ &
  $1.82\pm0.93$ &
  $\qquad$$p\bar{K}^{0}\eta^{0}$ &
  $1.18\pm0.44$
  \\
$\Lambda^{0}K^{+}\bar{K}^{0}$ &
 $0.73\pm0.37$ &
  $\qquad$$p\pi^{+}K^{-}$ &
  $28.3\pm2.6$ 
  \\
$\Sigma^{0}\pi^{+}\pi^{0}$ &
  $6.70\pm1.70$ &
  $\qquad$$n\pi^{+}\bar{K}^{0}$ &
  $4.80\pm1.76$ 
  \\
$\Sigma^{0}\pi^{+}\eta^{0}$ &
  $5.33\pm1.38$ &
  $\qquad$$\Sigma^{+}\pi^{0}\pi^{0}$ &
  $17.9\pm2.59$ 
  \\
$\Sigma^{0}K^{+}\bar{K}^{0}$ &
  $2.07\pm0.37$ &
  $\qquad$$\Sigma^{+}\pi^{0}\eta^{0}$ &
  $1.14\pm0.46$ 
  \\
$\Sigma^{-}\pi^{+}\pi^{+}$ &
  $6.91\pm2.27$ &
  $\qquad$$\Sigma^{+}\eta^{0}\eta^{0}$ &
  $0.18\pm0.07$
  \\
$\Xi^{-}\pi^{+}K^{+}$ &
  $0.95\pm0.28$ &
  $\qquad$$\Sigma^{+}\pi^{+}\pi^{-}$ &
  $6.53\pm1.59$ 
  \\
$\Xi^{0}\pi^{0}K^{+}$ &
  $1.94\pm0.54$ &
  $\qquad$$\Sigma^{+}K^{+}K^{-}$ &
  $0.37\pm0.15$ 
  \\
$\Xi^{0}K^{+}\eta^{0}$ &
  $0.27\pm0.04$ &
  $\qquad$$\Sigma^{+}K^{0}\bar{K}^{0}$ &
  $0.45\pm0.21$ 
  \\
$\Xi^{0}\pi^{+}K^{0}$ &
  $1.93\pm0.60$ &
  $\qquad$$\quad$ &
  $\quad$
  \\
  \hline
  \hline
DCS mode$\qquad$ &
  $10^{4}\mathcal{B}$ &
  $\qquad$DCS mode$\qquad$ &
  $10^{4}\mathcal{B}$
  \\ \hline
$\Lambda^{0}\pi^{0}K^{+}$ &
  $0.53\pm0.20$ &
  $\qquad$$p\pi^{0}\eta^{0}$ &
  $2.72\pm0.86$
  \\
$\Lambda^{0}K^{+}\eta^{0}$ &
  $(9.02\pm2.74)\times10^{-2}$ &
  $\qquad$$p\eta^{0}\eta^{0}$ &
  $(3.76\pm3.30)\times10^{-2}$
  \\
$\Lambda^{0}\pi^{+}K^{0}$ &
  $2.13\pm0.47$ &
  $\qquad$$p\pi^{+}\pi^{-}$ &
  $27.9\pm2.4$ 
  \\
$\Sigma^{0}\pi^{0}K^{+}$ &
  $3.31\pm0.33$ &
  $\qquad$$pK^{+}K^{-}$ &
  $1.11\pm0.19$ 
  \\
$\Sigma^{0}K^{+}\eta^{0}$ &
  $(4.58\pm1.05)\times10^{-2}$ &
  $\qquad$$pK^{0}\bar{K}^{0}$ &
  $0.14^{+0.19}_{-0.14}$ 

  \\
$\Sigma^{0}\pi^{+}K^{0}$ &
  $4.18\pm0.80$ &
  $\qquad$$n\pi^{+}\eta^{0}$ &
  $5.38\pm1.70$ 

  \\
$\Sigma^{-}\pi^{+}K^{+}$ &
  $0.69\pm0.21$ &
  $\qquad$$nK^{+}\bar{K}^{0}$ &
  $2.00\pm0.70$ 

  \\
$\Xi^{-}K^{+}K^{+}$ &
  $(2.45\pm1.09)\times10^{-2}$ &
  $\qquad$$\Sigma^{+}\pi^{0}K^{0}$ &
  $2.85\pm0.31$ 

  \\
$\Xi^{0}K^{+}K^{0}$ &
  $0.12\pm0.03$ &
  $\qquad$$\Sigma^{+}K^{0}\eta^{0}$&
  $(4.75\pm1.67)\times10^{-2}$ 

  \\
$p\pi^{0}\pi^{0}$ &
  $14.1\pm1.2$ &
  $\qquad$$\Sigma^{+}\pi^{-}K^{+}$ &
  $1.14\pm0.20$ 
  \\
  \hline
  
\end{tabular}

	\label{tab:Numerical result 2}
\end{table}

\begin{table}[H]
	\centering
	\caption{Numerical results for $\mathcal{B}(\Xi^{0}_{c}\rightarrow\mathbf{B}_{n}PP^{\prime})$}
	
\renewcommand{\arraystretch}{1}
\footnotesize
\begin{tabular}{llllll}
\hline
CF mode$\qquad$ &
  $10^{2}\mathcal{B}$ &
  $\qquad$CF mode$\qquad$ &
  $10^{2}\mathcal{B}$ 
  \\ \hline
$\Lambda^{0}\pi^{0}\bar{K}^{0}$&
  $1.18\pm0.30$&
  $\qquad$$\Xi^{0}\pi^{0}\eta^{0}$ &
  $0.33\pm0.09$ 

  \\
$\Lambda^{0}\bar{K}^{0}\eta^{0}$ &
  $0.28\pm0.05$ &
  $\qquad$$\Xi^{0}\eta^{0}\eta^{0}$ &
  $(7.50\pm5.18)\times10^{-3}$ 

  \\
$\Lambda^{0}\pi^{+}{K}^{-}$ &
  $1.25\pm0.30$ &
  $\qquad$$\Xi^{0}\pi^{+}\pi^{-}$&
  $0.28^{+0.38}_{-0.28}$ 

  \\
$\Sigma^{0}\pi^{0}\bar{K}^{0}$ &
  $2.26\pm0.51$ &
  $\qquad$$\Xi^{0}K^{+}K^{-}$ &
  $(9.57\pm0.77)\times10^{-2}$ 

  \\
$\Sigma^{0}\bar{K}^{0}\eta^{0}$ &
  $(9.61\pm1.61)\times10^{-2}$ &
  $\qquad$$\Xi^{0}K^{0}\bar{K}^{0}$ &
  $0.11\pm0.02$ 

  \\
$\Sigma^{0}\pi^{+}K^{-}$ &
  $2.85\pm0.61$ &
  $\qquad$$pK^{-}\bar{K}^{0}$ &
  $1.01\pm0.34 $ 

\\
$\Sigma^{-}\pi^{+}\bar{K}^{0}$ &
  $4.20\pm0.75$ &
  $\qquad$$n\bar{K}^{0}\bar{K}^{0}$ &
  $1.14\pm0.13$ 

  \\
$\Xi^{-}\pi^{+}\pi^{0}$ &
  $0.34\pm0.09$ &
  $\qquad$$\Sigma^{+}\pi^{0}K^{-}$ &
  $2.38\pm0.49$ 

  \\
$\Xi^{-}\pi^{+}\eta^{0}$ &
  $1.14\pm0.22$ &
  $\qquad$$\Sigma^{+}K^{-}\eta^{0}$ &
  $0.13\pm0.02$ 

  \\
$\Xi^{-}K^{+}\bar{K}^{0}$ &
  $0.43\pm0.14$ &
  $\qquad$$\Sigma^{+}\pi^{-}\bar{K}^{0}$ &
  $2.63\pm0.91$ 
  \\
$\Xi^{0}\pi^{0}\pi^{0}$ &
  $0.89\pm0.43$ &
  $\qquad$$\quad$&
  $\quad$
  \\ \hline
\hline
CS mode$\qquad$ &
  $10^{3}\mathcal{B}$ &
  $\qquad$CS mode$\qquad$ &
  $10^{3}\mathcal{B}$ 
  \\ \hline
$\Lambda^{0}\pi^{0}\pi^{0}$ &
 $2.25\pm0.66$&
  $\qquad$$\Xi^{-}\pi^{0}K^{+}$ &
  $1.99\pm0.26$ 

  \\
$\Lambda^{0}\pi^{0}\eta^{0}$ &
 $0.68\pm0.14$ &
  $\qquad$$\Xi^{-}K^{+}\eta^{0}$ &
  $(9.07\pm2.99)\times10^{-2}$ 

  \\
$\Lambda^{0}\eta^{0}\eta^{0}$ &
  $0.24\pm0.05$ &
  $\qquad$$\Xi^{-}\pi^{+}K^{0}$ &
  $1.33\pm0.22$

  \\
$\Lambda^{0}\pi^{+}\pi^{-}$ &
  $2.34\pm0.93$
&
  $\qquad$$\Xi^{0}\pi^{0}K^{0}$ &
  $0.63\pm0.10$ 

  \\
$\Lambda^{0}K^{+}K^{-}$ &
  $0.45\pm0.07$
&
  $\qquad$$\Xi^{0}K^{0}\eta^{0}$ &
  $(1.05\pm0.14)\times10^{-2}$ 
  \\
$\Lambda^{0}K^{0}\bar{K}^{0}$ &
 $0.38\pm0.05$ &
  $\qquad$$\Xi^{0}\pi^{-}K^{+}$ &
  $1.03\pm0.14$ 

\\
$\Sigma^{0}\pi^{0}\pi^{0}$ &
  $1.04\pm0.29$ 
&
  $\qquad$$p\pi^{0}K^{-}$ &
  $1.43\pm0.50$ 

  \\
$\Sigma^{0}\pi^{0}\eta^{0}$ &
  $1.53\pm0.27$
&
  $\qquad$$pK^{-}\eta^{0}$ &
  $0.30\pm0.13$ 

  \\
$\Sigma^{0}\eta^{0}\eta^{0}$ &
  $(3.00\pm1.24)\times10^{-2}$ &
  $\qquad$$p\pi^{-}\bar{K}^{0}$ &
  $5.78\pm1.89$ 

  \\
$\Sigma^{0}\pi^{+}\pi^{-}$ &
  $3.27\pm0.66$ 
&
  $\qquad$$n\pi^{0}\bar{K}^{0}$ &
  $8.89\pm1.73$ 
  \\
$\Sigma^{0}K^{+}K^{-}$ &
  $0.14\pm0.08$ 
&
  $\qquad$$n\bar{K}^{0}\eta^{0}$ &
  $0.27\pm0.10$ 

  \\
$\Sigma^{0}K^{0}\bar{K}^{0}$ &
  $(7.02\pm1.32)\times10^{-2}$ &
  $\qquad$$n\pi^{+}K^{-}$ &
  $8.69\pm1.65$ 

  \\
$\Sigma^{-}\pi^{+}\pi^{0}$ &
  $1.68\pm0.48$ 
&
  $\qquad$$\Sigma^{+}\pi^{0}\pi^{-}$ &
  $0.61\pm0.20$ 

  \\
$\Sigma^{-}\pi^{+}\eta^{0}$ &
  $6.16\pm0.77$ 
&
  $\qquad$$\Sigma^{+}\pi^{-}\eta^{0}$ &
  $0.68\pm0.30$ 

  \\
$\Sigma^{-}K^{+}\bar{K}^{0}$ &
  $0.41\pm0.07$ &
  $\qquad$$\Sigma^{+}K^{0}K^{-}$ &
  $0.59\pm0.11$ 
  \\
 \hline

\end{tabular}

	\label{tab:Numerical result 3.1}
\end{table}

\begin{table}[H]
	\centering
	\caption{(Continued) Numerical results for $\mathcal{B}(\Xi^{0}_{c}\rightarrow\mathbf{B}_{n}PP^{\prime})$}
	
\renewcommand{\arraystretch}{1}
\footnotesize
\begin{tabular}{llllll}
\hline
DCS mode$\qquad$ &
  $10^{5}\mathcal{B}$ &
  $\qquad$DCS mode$\qquad$ &
  $10^{5}\mathcal{B}$ &
  &
  \\ \hline
$\Lambda^{0}\pi^{0}K^{0}$ &
 $8.84\pm1.61$ 
&
  $\qquad$$\Xi^{0}K^{0}K^{0}$ &
  $(8.53\pm3.72)\times10^{-2}$ 

  \\
$\Lambda^{0}K^{0}\eta^{0}$ &
  $0.67\pm0.19$ 
&
  $\qquad$$p\pi^{-}\eta^{0}$ &
  $18.2\pm5.8$ 

  \\
$\Lambda^{0}\pi^{-}K^{+}$ &
  $13.7\pm2.7$ 
&
  $\qquad$$pK^{0}K^{-}$ &
  $6.79\pm2.37$ 

  \\
$\Sigma^{0}\pi^{0}K^{0}$ &
  $5.47\pm0.89$ 
&
  $\qquad$$n\pi^{0}\pi^{0}$ &
  $47.4\pm4.0$ 

  \\
$\Sigma^{0}K^{0}\eta^{0}$ &
  $(7.91\pm2.78)\times10^{-2}$ &
  $\qquad$$n\pi^{0}\eta^{0}$ &
  $9.14\pm2.90$ 
  \\
$\Sigma^{0}\pi^{-}K^{+}$ &
 $1.35\pm0.57$ 
&
  $\qquad$$n\eta^{0}\eta^{0}$ &
  $0.13\pm0.11$ 

\\
$\Sigma^{-}\pi^{0}K^{+}$ &
  $7.23\pm1.80$ 
&
  $\qquad$$n\pi^{+}\pi^{-}$ &
  $93.7\pm8.0$ 

  \\
$\Sigma^{-}K^{+}\eta^{0}$ &
  $0.30\pm0.07$ 
&
  $\qquad$$nK^{+}K^{-}$ &
  $0.48^{+0.65}_{-0.48}$ 

  \\
$\Sigma^{-}\pi^{+}K^{0}$ &
  $37.5\pm4.6$ 
&
  $\qquad$$nK^{0}\bar{K}^{0}$ &
  $3.63\pm0.61$ 

  \\
$\Xi^{-}K^{+}\bar{K}^{0}$ &
  $0.81\pm0.16$ &
  $\qquad$$\Sigma^{+}\pi^{-}K^{0}$ &
  $2.41\pm0.72$ 
  \\
  \hline
\end{tabular}

	\label{tab:Numerical result 3.2}
\end{table}

\begin{table}[H]
	\centering
	\caption{Numerical results for $\langle\alpha\rangle(\Lambda^{+}_{c}\rightarrow\mathbf{B}_{n}PP^{\prime})$}
	
\renewcommand{\arraystretch}{0.85}
\footnotesize
\begin{tabular}{llllll}
\hline
CF mode$\qquad$ &
  $\langle\alpha\rangle$ &
  $\qquad$CS mode$\qquad$ &
  $\langle\alpha\rangle$ &
  $\qquad$DCS mode$\qquad$ &
  $\langle\alpha\rangle$
  \\ \hline
$\Lambda^{0}\pi^{+}\eta^{0}$ &
  $0.1\pm0.6$&
  $\qquad$$\Lambda^{0}\pi^{0}K^{+}$ &
   $-0.4\pm0.4$&
  $\qquad$$\Lambda^{0}K^{+}K^{0}$ &
  $0.14\pm0.30$\\
  
$\Lambda^{0}K^{+}\bar{K}^{0}$ &
   $0.63\pm0.30$&
  $\qquad$$\Lambda^{0}K^{+}\eta^{0}$ &
   $0.17\pm0.32$&
  $\qquad$$\Sigma^{0}K^{+}K^{0}$ &
  $-0.3\pm0.4$\\
  
$\Sigma^{0}\pi^{+}\pi^{0}$ &
  $0.73^{+0.27}_{-0.35}$&
  $\qquad$$\Lambda^{0}\pi^{+}K^{0}$ &
   $-0.96\pm0.04$&
  $\qquad$$\Sigma^{-}K^{+}K^{+}$ &
  $-0.3\pm0.4$\\
  
$\Sigma^{0}\pi^{+}\eta^{0}$ &
   $-0.86^{+0.28}_{-0.14}$&
  $\qquad$$\Sigma^{0}\pi^{0}K^{+}$ &
   $0.965\pm0.021$&
  $\qquad$$p\pi^{0}K^{0}$ &
   $-0.965^{+0.05}_{-0.035}$\\
   
$\Sigma^{0}K^{+}\bar{K}^{0}$ &
  $-0.80^{+0.5}_{-0.20}$&
  $\qquad$$\Sigma^{0}K^{+}\eta^{0}$ &
   $0.15\pm0.21$&
  $\qquad$$pK^{0}\eta^{0}$ &
   $-0.74^{+0.4}_{-0.26}$\\
   
$\Sigma^{-}\pi^{+}\pi^{+}$ &
  $0.73^{+0.27}_{-0.35}$&
  $\qquad$$\Sigma^{0}\pi^{+}K^{0}$ &
  $0.3\pm0.7$&
  $\qquad$$p\pi^{-}K^{+}$ &
  $0.69\pm0.27$\\
  
$\Xi^{-}\pi^{+}K^{+}$ &
  $-0.85^{+0.4}_{-0.15}$&
  $\qquad$$\Sigma^{-}\pi^{+}K^{+}$ &
  $0.6\pm0.4$&
  $\qquad$$n\pi^{0}K^{+}$ &
   $-0.93^{+0.10}_{-0.07}$\\
   
$\Xi^{0}\pi^{0}K^{+}$ &
  $-0.54\pm0.33$&
  $\qquad$$\Xi^{-}K^{+}K^{+}$ &
  $-0.75^{+0.5}_{-0.25}$&
  $\qquad$$nK^{+}\eta^{0}$ &
   $0.6^{+0.4}_{-0.7}$\\
   
$\Xi^{0}K^{+}\eta^{0}$ &
  $-0.29\pm0.20$&
  $\qquad$$\Xi^{0}K^{+}K^{0}$ &
  $-0.72^{+0.5}_{-0.28}$&
  $\qquad$$n\pi^{+}K^{0}$ &
  $0.60\pm0.31$\\
  
$\Xi^{0}\pi^{+}K^{0}$ &
  $-0.04\pm0.35$&
  $\qquad$$p\pi^{0}\pi^{0}$ &
   $0.86^{+0.14}_{-0.25}$&
  $\qquad$$\Sigma^{+}K^{0}K^{0}$ &
  $-0.3\pm0.4$\\
  
$p\pi^{0}\bar{K}^{0}$ &
   $0.6^{+0.4}_{-0.5}$&
  $\qquad$$p\pi^{0}\eta^{0}$ &
   $-0.93^{+0.16}_{-0.07}$&
  $\qquad$$\quad$ &
  $\quad$ 
  \\
  
$p\bar{K}^{0}\eta^{0}$ &
   $0.0\pm0.5$&
  $\qquad$$p\eta^{0}\eta^{0}$ &
   $-0.6\pm0.4$&
  $\qquad$$\quad$ &
  $\quad$ 
  \\
  
$p\pi^{+}K^{-}$ &
  $0.80^{+0.20}_{-0.4}$&
  $\qquad$$p\pi^{+}\pi^{-}$ &
   $-0.65^{+0.4}_{-0.35}$&
  $\qquad$$\quad$ &
  $\quad$ 
  \\
  
$n\pi^{+}\bar{K}^{0}$ &
  $-0.68^{+0.5}_{-0.32}$&
  $\qquad$$pK^{+}K^{-}$ &
   $-0.88^{+0.20}_{-0.12}$&
  $\qquad$$\quad$ &
  $\quad$ 
  \\
  
$\Sigma^{+}\pi^{0}\pi^{0}$ &
   $-0.95^{+0.10}_{-0.05}$&
  $\qquad$$pK^{0}\bar{K}^{0}$ &
  $-0.3\pm0.7$&
  $\qquad$$\quad$ &
  $\quad$ 
  \\
  
$\Sigma^{+}\pi^{0}\eta^{0}$ &
   $-0.86^{+0.28}_{-0.14}$&
  $\qquad$$n\pi^{+}\pi^{0}$ &
  $0.3\pm0.7$&
  $\qquad$$\quad$ &
  $\quad$ 
  \\
  
$\Sigma^{+}\eta^{0}\eta^{0}$ &
   $0.10\pm0.07$&
  $\qquad$$n\pi^{+}\eta^{0}$ &
   $0.6^{+0.4}_{-0.5}$&
  $\qquad$$\quad$ &
  $\quad$ 
  \\
  
$\Sigma^{+}\pi^{+}\pi^{-}$ &
   $-0.1\pm0.5$&
  $\qquad$$nK^{+}\bar{K}^{0}$ &
   $0.89^{+0.11}_{-0.27}$&
  $\qquad$$\quad$ &
  $\quad$ 
  \\
  
$\Sigma^{+}K^{+}K^{-}$ &
  $0.0\pm0.9$&
  $\qquad$$\Sigma^{+}\pi^{0}K^{0}$ &
  $0.3\pm0.7$&
  $\qquad$$\quad$ &
  $\quad$ 
  \\
  
$\Sigma^{+}K^{0}\bar{K}^{0}$ &
  $0.0\pm0.4$&
  $\qquad$$\Sigma^{+}K^{0}\eta^{0}$ &
   $0.15\pm0.21$&
  $\qquad$$\quad$ &
  $\quad$ 
  \\
  
$\quad$ &
  $\quad$ &
  $\qquad$$\Sigma^{+}\pi^{-}K^{+}$ &
  $0.91^{+0.09}_{-0.21}$&
  $\qquad$$\quad$ &
  $\quad$ 
  \\ 
  \hline
\end{tabular}

	\label{tab:Numerical result 4}
\end{table}
We note that the $\chi^2$ fit suffers a $Z_2$ ambiguity of ${\cal B}\,(A,B) = {\cal B}\,(A,-B)$. The ambiguity can be broken with an input of $\langle \alpha \rangle$ as 
$\langle \alpha \rangle\,(A,B) =  \langle \alpha \rangle \,(A,-B).$ Here, we choose the  $\langle\alpha\rangle$ of $\Lambda^{+}_{c}\rightarrow\Xi^{0}\pi^{0}K^{+}$ and $\Lambda^{+}_{c}\rightarrow\Xi^{0}K^{+}\eta^{0}$ to be negative as their amplitudes are  contributed by  Fig.~\ref{fig2} mainly, where $x$ and $y$ have a negative helicity in the chiral limit. Consequently, the helicity of $\mathbf{B}_{n}$ is negative also, leading to a negative averaged up-down asymmetry. We list the predictions for the up-down asymmetries of $\langle\alpha\rangle\,(\Lambda^{+}_{c},\Xi^{+}_{c},\Xi^{0}_{c}\rightarrow\mathbf{B}_{n}PP^{\prime})$ in Tables \ref{tab:Numerical result 4}, \ref{tab:Numerical result 5} and \ref{tab:Numerical result 6}. Meanwhile, without considering CP violation of physical particles $K^{0}_{S}$ and $K^{0}_{L}$, we can also give branching ratios and up-down asymmetries of three-body decay channels involving $K^{0}_{S}$ and $K^{0}_{L}$ of mixed-modes, which are presented in Table \ref{tab:Numerical result 7}. These results are acquired under the assumption of S-wave meson pairs in the final states, neglecting the contributions from pseudoscalar meson exchanges, which can be used to assess the dominance of S-wave meson pairs in nonleptonic three-body decays.
\begin{table}[H]
    \centering
    \caption{Numerical results for $\langle\alpha\rangle(\Xi^{+}_{c}\rightarrow\mathbf{B}_{n}PP^{\prime})$}
    
    \renewcommand{\arraystretch}{0.85}
\footnotesize
\begin{tabular}{llllll}
\hline
CF mode$\qquad$ &
  $\langle\alpha\rangle$ &
  $\qquad$CS mode$\qquad$ &
  $\langle\alpha\rangle$ &
  $\qquad$DCS mode$\qquad$ &
  $\langle\alpha\rangle$
  \\ \hline
$\Lambda^{0}\pi^{+}\bar{K}^{0}$ &
  $-0.5^{+0.8}_{-0.5}$&
  $\qquad$$\Lambda^{0}\pi^{+}\pi^{0}$ &
  $0.3\pm0.6$&
  $\qquad$$\Lambda^{0}\pi^{0}K^{+}$ &
  $-0.90^{+0.24}_{-0.10}$\\
$\Sigma^{0}\pi^{+}\bar{K}^{0}$ &
  $0.0\pm0.6$&
  $\qquad$$\Lambda^{0}\pi^{+}\eta^{0}$ &
  $0.5^{+0.5}_{-0.6}$&
  $\qquad$$\Lambda^{0}K^{+}\eta^{0}$ &
  $0.78^{+0.22}_{-0.33}$\\
$\Xi^{-}\pi^{+}\pi^{+}$ &
  $-0.6^{+0.6}_{-0.4}$&
  $\qquad$$\Lambda^{0}K^{+}\bar{K}^{0}$ &
  $-0.2\pm0.4$&
  $\qquad$$\Lambda^{0}\pi^{+}K^{0}$ &
  $-0.85^{+0.33}_{-0.15}$\\
$\Xi^{0}\pi^{+}\pi^{0}$ &
  $-0.6^{+0.6}_{-0.4}$&
  $\qquad$$\Sigma^{0}\pi^{+}\pi^{0}$ &
  $-0.94^{+0.18}_{-0.06}$&
  $\qquad$$\Sigma^{0}\pi^{0}K^{+}$ &
  $-0.1\pm0.5$\\
$\Xi^{0}\pi^{+}\eta^{0}$ &
  $-0.96^{+0.04}_{-0.04}$&
  $\qquad$$\Sigma^{0}\pi^{+}\eta^{0}$ &
  $0.77^{+0.23}_{-0.24}$&
  $\qquad$$\Sigma^{0}K^{+}\eta^{0}$ &
  $-0.73^{+0.5}_{-0.27}$\\
$\Xi^{0}K^{+}\bar{K}^{0}$ &
  $0.95^{+0.05}_{-0.10}$&
  $\qquad$$\Sigma^{0}K^{+}\bar{K}^{0}$ &
  $-0.96^{+0.06}_{-0.04}$&
  $\qquad$$\Sigma^{0}\pi^{+}K^{0}$ &
  $0.95^{+0.05}_{-0.08}$\\
$p\bar{K}^{0}\bar{K}^{0}$ &
  $-0.67^{+0.7}_{-0.33}$&
  $\qquad$$\Sigma^{-}\pi^{+}\pi^{+}$ &
  $-0.6^{+0.5}_{-0.4}$&
  $\qquad$$\Sigma^{-}\pi^{+}K^{+}$ &
  $-0.66^{+0.5}_{-0.34}$\\
$\Sigma^{+}\pi^{0}\bar{K}^{0}$ &
  $-0.85^{+0.30}_{-0.15}$&
  $\qquad$$\Xi^{-}\pi^{+}K^{+}$ &
  $0.6\pm0.4$&
  $\qquad$$\Xi^{-}K^{+}K^{+}$ &
  $0.96^{+0.04}_{-0.05}$\\
$\Sigma^{+}\bar{K}^{0}\eta^{0}$ &
  $-0.95^{+0.11}_{-0.05}$&
  $\qquad$$\Xi^{0}\pi^{0}K^{+}$ &
  $0.5\pm0.4$&
  $\qquad$$\Xi^{0}K^{+}K^{0}$ &
  $-0.32\pm0.29$\\
$\Sigma^{+}\pi^{+}K^{-}$ &
  $0.79^{+0.21}_{-0.26}$&
  $\qquad$$\Xi^{0}K^{+}\eta^{0}$ &
  $0.16\pm0.33$&
  $\qquad$$p\pi^{0}\pi^{0}$ &
  $0.0\pm0.4$\\
$\quad$ &
  $\quad$ &
  $\qquad$$\Xi^{0}\pi^{+}K^{0}$ &
  $0.84^{+0.16}_{-0.33}$&
  $\qquad$$p\pi^{0}\eta^{0}$ &
  $-0.87^{+0.28}_{-0.13}$\\
$\quad$ &
  $\quad$ &
  $\qquad$$p\pi^{0}\bar{K}^{0}$ &
  $0.969\pm0.023$&
  $\qquad$$p\eta^{0}\eta^{0}$ &
  $-0.6^{+0.9}_{-0.4}$\\
$\quad$ &
  $\quad$ &
  $\qquad$$p\bar{K}^{0}\eta^{0}$ &
  $-0.74^{+0.5}_{-0.26}$&
  $\qquad$$p\pi^{+}\pi^{-}$ &
  $0.0\pm0.4$\\
$\quad$ &
  $\quad$ &
  $\qquad$$p\pi^{+}K^{-}$ &
  $0.70^{+0.30}_{-0.32}$&
  $\qquad$$pK^{+}K^{-}$ &
  $-0.1\pm0.6$\\
$\quad$ &
  $\quad$ &
  $\qquad$$n\pi^{+}\bar{K}^{0}$ &
  $-0.5\pm0.4$&
  $\qquad$$pK^{0}\bar{K}^{0}$ &
  $0.1\pm0.9$\\
$\quad$ &
  $\quad$ &
  $\qquad$$\Sigma^{+}\pi^{0}\pi^{0}$ &
  $-0.94^{+0.16}_{-0.06}$&
  $\qquad$$n\pi^{+}\eta^{0}$ &
  $-0.87^{+0.28}_{-0.13}$\\
$\quad$ &
  $\quad$ &
  $\qquad$$\Sigma^{+}\pi^{0}\eta^{0}$ &
  $0.94^{+0.06}_{-0.13}$&
  $\qquad$$nK^{+}\bar{K}^{0}$ &
  $0.0\pm0.6$\\
$\quad$ &
  $\quad$ &
  $\qquad$$\Sigma^{+}\eta^{0}\eta^{0}$ &
  $0.3\pm0.4$&
  $\qquad$$\Sigma^{+}\pi^{0}K^{0}$ &
  $0.2\pm0.6$\\
$\quad$ &
  $\quad$ &
  $\qquad$$\Sigma^{+}\pi^{+}\pi^{-}$ &
  $-0.81^{+0.26}_{-0.19}$&
  $\qquad$$\Sigma^{+}K^{0}\eta^{0}$&
  $0.64\pm0.32$\\
$\quad$ &
  $\quad$ &
  $\qquad$$\Sigma^{+}K^{+}K^{-}$ &
  $-0.92^{+0.25}_{-0.08}$&
  $\qquad$$\Sigma^{+}\pi^{-}K^{+}$ &
  $0.89^{+0.11}_{-0.29}$\\
$\quad$ &
  $\quad$ &
  $\qquad$$\Sigma^{+}K^{0}\bar{K}^{0}$ &
  $-0.3\pm0.5$&
  $\qquad$$\quad$ &
  $\quad$ \\ \hline
\end{tabular}

    \label{tab:Numerical result 5}
\end{table}

\begin{table}[H]
    \centering
    \caption{Numerical results for $\langle\alpha\rangle(\Xi^{0}_{c}\rightarrow\mathbf{B}_{n}PP^{\prime})$}
    
\renewcommand{\arraystretch}{1}
\footnotesize
\begin{tabular}{llllll}
\hline
CF mode$\qquad$ &
  $\langle\alpha\rangle$ &
  $\qquad$CS mode$\qquad$ &
  $\langle\alpha\rangle$ &
  $\qquad$DCS mode$\qquad$ &
  $\langle\alpha\rangle$
  \\ \hline
$\Lambda^{0}\pi^{0}\bar{K}^{0}$&
 $0.1\pm0.5$&
  $\qquad$$\Lambda^{0}\pi^{0}\pi^{0}$ &
  $0.965^{+0.035}_{-0.08}$&
  $\qquad$$\Lambda^{0}\pi^{0}K^{0}$ &
  $-0.95^{+0.09}_{-0.05}$\\
$\Lambda^{0}\bar{K}^{0}\eta^{0}$ &
  $0.6^{+0.4}_{-0.5}$&
  $\qquad$$\Lambda^{0}\pi^{0}\eta^{0}$ &
  $0.967\pm0.016$&
  $\qquad$$\Lambda^{0}K^{0}\eta^{0}$ &
  $0.0\pm0.5$\\
$\Lambda^{0}\pi^{+}{K}^{-}$ &
  $0.4\pm0.5$&
  $\qquad$$\Lambda^{0}\eta^{0}\eta^{0}$ &
  $0.31\pm0.35$&
  $\qquad$$\Lambda^{0}\pi^{-}K^{+}$ &
  $-0.71^{+0.31}_{-0.29}$\\
$\Sigma^{0}\pi^{0}\bar{K}^{0}$ &
  $-0.1\pm0.5$&
  $\qquad$$\Lambda^{0}\pi^{+}\pi^{-}$ &
  $0.88^{+0.12}_{-0.34}$&
  $\qquad$$\Sigma^{0}\pi^{0}K^{0}$ &
  $0.965\pm0.034$\\
$\Sigma^{0}\bar{K}^{0}\eta^{0}$ &
  $-0.67\pm0.29$&
  $\qquad$$\Lambda^{0}K^{+}K^{-}$ &
  $-0.3\pm0.7$&
  $\qquad$$\Sigma^{0}K^{0}\eta^{0}$ &
  $0.64\pm0.32$\\
$\Sigma^{0}\pi^{+}K^{-}$ &
 $-0.93^{+0.14}_{-0.07}$&
  $\qquad$$\Lambda^{0}K^{0}\bar{K}^{0}$ &
  $-0.89^{+0.18}_{-0.11}$&
  $\qquad$$\Sigma^{0}\pi^{-}K^{+}$ &
$0.2\pm0.7$\\
$\Sigma^{-}\pi^{+}\bar{K}^{0}$ &
  $0.71^{+0.29}_{-0.33}$&
  $\qquad$$\Sigma^{0}\pi^{0}\pi^{0}$ &
  $0.0\pm0.6$&
  $\qquad$$\Sigma^{-}\pi^{0}K^{+}$ &
  $0.4\pm0.5$\\
$\Xi^{-}\pi^{+}\pi^{0}$ &
  $-0.6^{+0.6}_{-0.4}$&
  $\qquad$$\Sigma^{0}\pi^{0}\eta^{0}$ &
  $0.68^{+0.32}_{-0.4}$&
  $\qquad$$\Sigma^{-}K^{+}\eta^{0}$ &
  $-0.74^{+0.5}_{-0.26}$\\
$\Xi^{-}\pi^{+}\eta^{0}$ &
  $0.1\pm0.6$&
  $\qquad$$\Sigma^{0}\eta^{0}\eta^{0}$ &
  $0.3\pm0.4$&
  $\qquad$$\Sigma^{-}\pi^{+}K^{0}$ &
  $0.80^{+0.20}_{-0.26}$\\
$\Xi^{-}K^{+}\bar{K}^{0}$ &
  $0.961\pm0.028$&
  $\qquad$$\Sigma^{0}\pi^{+}\pi^{-}$ &
  $-0.73^{+0.33}_{-0.27}$&
  $\qquad$$\Xi^{-}K^{+}\bar{K}^{0}$ &
  $0.40\pm0.25$\\
$\Xi^{0}\pi^{0}\pi^{0}$ &
  $-0.3\pm0.7$&
  $\qquad$$\Sigma^{0}K^{+}K^{-}$ &
   $0.1\pm0.5$&
  $\qquad$$\Xi^{0}K^{0}K^{0}$ &
  $0.96^{+0.04}_{-0.06}$\\
$\Xi^{0}\pi^{0}\eta^{0}$ &
  $0.94^{+0.06}_{-0.14}$&
  $\qquad$$\Sigma^{0}K^{0}\bar{K}^{0}$ &
  $-0.83^{+0.23}_{-0.17}$&
  $\qquad$$p\pi^{-}\eta^{0}$ &
  $-0.87^{+0.28}_{-0.13}$\\
$\Xi^{0}\eta^{0}\eta^{0}$ &
  $-0.86^{+0.27}_{-0.14}$&
  $\qquad$$\Sigma^{-}\pi^{+}\pi^{0}$ &
  $0.97\pm0.02$&
  $\qquad$$pK^{0}K^{-}$ &
  $0.0\pm0.6$\\
$\Xi^{0}\pi^{+}\pi^{-}$&
  $0.0\pm0.9$&
  $\qquad$$\Sigma^{-}\pi^{+}\eta^{0}$ &
  $0.96^{+0.04}_{-0.07}$&
  $\qquad$$n\pi^{0}\pi^{0}$ &
  $0.0\pm0.4$\\
$\Xi^{0}K^{+}K^{-}$ &
  $0.0\pm0.9$&
  $\qquad$$\Sigma^{-}K^{+}\bar{K}^{0}$ &
  $-0.2\pm0.7$&
  $\qquad$$n\pi^{0}\eta^{0}$ &
  $-0.87^{+0.28}_{-0.13}$\\
$\Xi^{0}K^{0}\bar{K}^{0}$ &
  $-0.06\pm0.24$&
  $\qquad$$\Xi^{-}\pi^{0}K^{+}$ &
  $0.96\pm0.04$&
  $\qquad$$n\eta^{0}\eta^{0}$ &
  $-0.6^{+0.9}_{-0.4}$\\
$pK^{-}\bar{K}^{0}$ &
  $-0.6^{+0.5}_{-0.4}$&
  $\qquad$$\Xi^{-}K^{+}\eta^{0}$ &
  $0.71\pm0.22$&
  $\qquad$$n\pi^{+}\pi^{-}$ &
  $0.0\pm0.4$\\
$n\bar{K}^{0}\bar{K}^{0}$ &
  $0.67^{+0.33}_{-0.35}$&
  $\qquad$$\Xi^{-}\pi^{+}K^{0}$ &
  $-0.2\pm0.6$&
  $\qquad$$nK^{+}K^{-}$ &
  $0.1\pm0.9$\\
$\Sigma^{+}\pi^{0}K^{-}$ &
  $-0.74^{+0.35}_{-0.26}$&
  $\qquad$$\Xi^{0}\pi^{0}K^{0}$ &
  $-0.965\pm0.025$&
  $\qquad$$nK^{0}\bar{K}^{0}$ &
  $-0.1\pm0.6$\\
$\Sigma^{+}K^{-}\eta^{0}$ &
  $-0.54\pm0.33$&
  $\qquad$$\Xi^{0}K^{0}\eta^{0}$ &
  $-0.2^{+1.1}_{-0.8}$&
  $\qquad$$\Sigma^{+}\pi^{-}K^{0}$ &
  $-0.65^{+0.5}_{-0.35}$\\
$\Sigma^{+}\pi^{-}\bar{K}^{0}$ &
  $0.0\pm0.5$&
  $\qquad$$\Xi^{0}\pi^{-}K^{+}$ &
  $-0.5\pm0.5$&
  $\qquad$$\quad$ &
  $\quad$ 
  \\
$\quad$ &
  $\quad$ &
  $\qquad$$p\pi^{0}K^{-}$ &
  $0.0\pm0.7$&
  $\qquad$$\quad$ &
  $\quad$ 
  \\
$\quad$ &
  $\quad$ &
  $\qquad$$pK^{-}\eta^{0}$ &
  $0.78^{+0.22}_{-0.4}$&
  $\qquad$$\quad$ &
  $\quad$ 
  \\
$\quad$ &
  $\quad$ &
  $\qquad$$p\pi^{-}\bar{K}^{0}$ &
  $-0.88^{+0.27}_{-0.12}$&
  $\qquad$$\quad$ &
  $\quad$ 
  \\
$\quad$ &
  $\quad$ &
  $\qquad$$n\pi^{0}\bar{K}^{0}$ &
  $-0.4\pm0.4$&
  $\qquad$$\quad$ &
  $\quad$ 
  \\
$\quad$ &
  $\quad$ &
  $\qquad$$n\bar{K}^{0}\eta^{0}$ &
  $-0.3\pm0.7$&
  $\qquad$$\quad$ &
  $\quad$ 
  \\
$\quad$ &
  $\quad$ &
  $\qquad$$n\pi^{+}K^{-}$ &
  $-0.31\pm0.35$&
  $\qquad$$\quad$ &
  $\quad$ 
  \\
$\quad$ &
  $\quad$ &
  $\qquad$$\Sigma^{+}\pi^{0}\pi^{-}$ &
  $-0.6^{+0.5}_{-0.4}$&
  $\qquad$$\quad$ &
  $\quad$ 
  \\
$\quad$ &
  $\quad$ &
  $\qquad$$\Sigma^{+}\pi^{-}\eta^{0}$ &
  $0.5\pm0.5$&
  $\qquad$$\quad$ &
  $\quad$ 
  \\
$\quad$ &
  $\quad$ &
  $\qquad$$\Sigma^{+}K^{0}K^{-}$ &
    $-0.58\pm0.35$&
  $\qquad$$\quad$ &
  $\quad$ 
  \\ \hline
  \end{tabular}

    \label{tab:Numerical result 6}
\end{table}

\begin{table}[H]
    \centering
    \caption{Decay branching ratios and averaged up-down asymmetries for CF and DCS mixed-mode processes involving $K^{0}_{S}$ and $K^{0}_{L}$}
    
\footnotesize
\renewcommand{\arraystretch}{1}
\begin{tabular}{llllll}
\hline
Channels&$\quad$$\mathcal{B}$&$\quad$$\langle\alpha\rangle$&$\quad$Channels&$\quad$$\mathcal{B}$&$\quad$$\langle\alpha\rangle$\\
\hline
$\Lambda^{+}_{c}\rightarrow\Lambda^{0}K^{+}K^{0}_{S}$&
               $\quad$$(2.73\pm0.49)\times10^{-3}$&
               $\quad$$0.65\pm0.30$&
               $\quad$$\Xi^{0}_{c}\rightarrow\Sigma^{+}\pi^{-}K^{0}_{S}$&
               $\quad$$(1.35\pm0.45)\times10^{-2}$&
               $\quad$$-0.10\pm0.50$\\

               $\Lambda^{+}_{c}\rightarrow\Lambda^{0}K^{+}K^{0}_{L}$&
               $\quad$$(3.13\pm0.53)\times10^{-3}$&
               $\quad$$0.62\pm0.30$&
               $\quad$$\Xi^{0}_{c}\rightarrow\Sigma^{+}\pi^{-}K^{0}_{L}$&
               $\quad$$(1.29\pm0.46)\times10^{-2}$&
               $\quad$$0.00\pm0.50$\\

               $\Lambda^{+}_{c}\rightarrow\Sigma^{0}K^{+}K^{0}_{S}$&
               $\quad$$(1.27\pm0.48)\times10^{-4}$&
               $\quad$$-0.85^{+0.40}_{-0.15}$&
               $\quad$$\Xi^{0}_{c}\rightarrow\Sigma^{0}\pi^{0}K^{0}_{S}$&
               $\quad$$(1.06\pm0.24)\times10^{-2}$&
               $\quad$$-0.20\pm0.50$\\

               $\Lambda^{+}_{c}\rightarrow\Sigma^{0}K^{+}K^{0}_{L}$&
               $\quad$$(1.08\pm0.35)\times10^{-4}$&
               $\quad$$-0.73^{+0.60}_{-0.27}$&
               $\quad$$\Xi^{0}_{c}\rightarrow\Sigma^{0}\pi^{0}K^{0}_{L}$&
               $\quad$$(1.20\pm0.27)\times10^{-2}$&
               $\quad$$0.00\pm0.50$\\

               $\Lambda^{+}_{c}\rightarrow p\pi^{0}K^{0}_{S}$&
               $\quad$$(1.88\pm0.14)\times10^{-2}$&
               $\quad$$0.60^{+0.40}_{-0.50}$&
               $\quad$$\Xi^{0}_{c}\rightarrow\Sigma^{-}\pi^{+}K^{0}_{S}$&
               $\quad$$(1.81\pm0.34)\times10^{-2}$&
               $\quad$$0.60\pm0.40$\\

               $\Lambda^{+}_{c}\rightarrow p\pi^{0}K^{0}_{L}$&
               $\quad$$(2.02\pm0.15)\times10^{-2}$&
               $\quad$$0.50\pm0.50$&
               $\quad$$\Xi^{0}_{c}\rightarrow\Sigma^{-}\pi^{+}K^{0}_{L}$&
               $\quad$$(2.43\pm0.41)\times10^{-2}$&
               $\quad$$0.78^{+0.22}_{-0.28}$\\

               $\Lambda^{+}_{c}\rightarrow p\eta^{0}K^{0}_{S}$&
               $\quad$$(3.41\pm0.49)\times10^{-3}$&
               $\quad$$0.00\pm0.50$&
               $\quad$$\Xi^{0}_{c}\rightarrow\Xi^{-}K^{+}K^{0}_{S}$&
               $\quad$$(1.97\pm0.65)\times10^{-3}$&
               $\quad$$0.96\pm0.00$\\

               $\Lambda^{+}_{c}\rightarrow p\eta^{0}K^{0}_{L}$&
               $\quad$$(3.97\pm0.58)\times10^{-3}$&
               $\quad$$0.00\pm0.50$&
               $\quad$$\Xi^{0}_{c}\rightarrow\Xi^{-}K^{+}K^{0}_{L}$&
               $\quad$$(2.29\pm0.75)\times10^{-3}$&
               $\quad$$0.96^{+0.04}_{-0.05}$\\

               $\Lambda^{+}_{c}\rightarrow n\pi^{+}K^{0}_{S}$&
               $\quad$$(1.85\pm0.09)\times10^{-2}$&
               $\quad$$-0.76^{+0.40}_{-0.24}$&
               $\quad$$\Xi^{0}_{c}\rightarrow pK^{-}K^{0}_{S}$&
               $\quad$$(5.37\pm1.70)\times10^{-3}$&
               $\quad$$-0.69^{+0.40}_{-0.31}$\\

               $\Lambda^{+}_{c}\rightarrow n\pi^{+}K^{0}_{L}$&
               $\quad$$(1.83\pm0.17)\times10^{-2}$&
               $\quad$$-0.60^{+0.50}_{-0.40}$&
               $\quad$$\Xi^{0}_{c}\rightarrow pK^{-}K^{0}_{L}$&
               $\quad$$(4.79\pm1.67)\times10^{-3}$&
               $\quad$$-0.50\pm0.50$\\

               $\Lambda^{+}_{c}\rightarrow\Sigma^{+}K^{0}_{S}K^{0}_{S}$&
               $\quad$$(1.00^{+1.26}_{-1.00})\times10^{-4}$&
               $\quad$$0.00\pm0.40$&
               $\quad$$\Xi^{0}_{c}\rightarrow\Lambda^{0}\eta^{0}K^{0}_{S}$&
               $\quad$$(1.37\pm0.24)\times10^{-3}$&
               $\quad$$0.60^{+0.40}_{-0.50}$\\

               $\Lambda^{+}_{c}\rightarrow\Sigma^{+}K^{0}_{S}K^{0}_{L}$&
               $\quad$$(5.46\pm2.09)\times10^{-7}$&
               $\quad$$-0.30\pm0.40$&
               $\quad$$\Xi^{0}_{c}\rightarrow\Lambda^{0}\eta^{0}K^{0}_{L}$&
               $\quad$$(1.47\pm0.26)\times10^{-3}$&
               $\quad$$0.69^{+0.31}_{-0.40}$\\

               $\Lambda^{+}_{c}\rightarrow\Sigma^{+}K^{0}_{L}K^{0}_{L}$&
               $\quad$$(0.73^{+1.10}_{-0.73})\times10^{-4}$&
               $\quad$$0.00\pm0.50$&
               $\quad$$\Xi^{0}_{c}\rightarrow\Xi^{0}K^{0}_{S}K^{0}_{S}$&
               $\quad$$(1.10\pm0.19)\times10^{-3}$&
               $\quad$$-0.09\pm0.24$\\

               $\Xi^{+}_{c}\rightarrow\Sigma^{+}\pi^{0}K^{0}_{S}$&
               $\quad$$(2.67\pm0.50)\times10^{-2}$&
               $\quad$$-0.89^{+0.25}_{-0.11}$&
               $\quad$$\Xi^{0}_{c}\rightarrow\Xi^{0}K^{0}_{S}K^{0}_{L}$&
               $\quad$$(4.27\pm1.86)\times10^{-7}$&
               $\quad$$0.96^{+0.04}_{-0.06}$\\

               $\Xi^{+}_{c}\rightarrow\Sigma^{+}\pi^{0}K^{0}_{L}$&
               $\quad$$(3.32\pm0.59)\times10^{-2}$&
               $\quad$$-0.81^{+0.35}_{-0.19}$&
               $\quad$$\Xi^{0}_{c}\rightarrow\Xi^{0}K^{0}_{L}K^{0}_{L}$&
               $\quad$$(1.15\pm0.20)\times10^{-3}$&
               $\quad$$-0.03\pm0.24$\\

               $\Xi^{+}_{c}\rightarrow\Sigma^{+}\eta^{0}K^{0}_{S}$&
               $\quad$$(1.35\pm0.80)\times10^{-4}$&
               $\quad$$-0.96^{+0.07}_{-0.04}$&
               $\quad$$\Xi^{0}_{c}\rightarrow nK^{0}_{S}K^{0}_{S}$&
               $\quad$$(5.44\pm0.63)\times10^{-3}$&
               $\quad$$0.60\pm0.40$\\

               $\Xi^{+}_{c}\rightarrow\Sigma^{+}\eta^{0}K^{0}_{L}$&
               $\quad$$(1.71\pm0.99)\times10^{-4}$&
               $\quad$$-0.90^{+0.24}_{-0.10}$&
               $\quad$$\Xi^{0}_{c}\rightarrow nK^{0}_{S}K^{0}_{L}$&
               $\quad$$(5.69\pm0.64)\times10^{-3}$&
               $\quad$$0.67^{+0.33}_{-0.35}$\\

               $\Xi^{+}_{c}\rightarrow\Sigma^{0}\pi^{+}K^{0}_{S}$&
               $\quad$$(3.33\pm0.55)\times10^{-2}$&
               $\quad$$-0.10\pm0.60$&
               $\quad$$\Xi^{0}_{c}\rightarrow nK^{0}_{L}K^{0}_{L}$&
               $\quad$$(6.00\pm0.68)\times10^{-3}$&
               $\quad$$0.76^{+0.24}_{-0.31}$\\

               $\Xi^{+}_{c}\rightarrow\Sigma^{0}\pi^{+}K^{0}_{L}$&
               $\quad$$(4.04\pm0.67)\times10^{-2}$&
               $\quad$$0.00\pm0.60$&
               $\quad$$\Xi^{0}_{c}\rightarrow\Lambda^{0}\pi^{0}K^{0}_{S}$&
               $\quad$$(6.65\pm1.64)\times10^{-3}$&
               $\quad$$0.00\pm0.50$\\

               $\Xi^{+}_{c}\rightarrow\Xi^{0}K^{+}K^{0}_{S}$&
               $\quad$$(8.32\pm3.76)\times10^{-4}$&
               $\quad$$0.92^{+0.08}_{-0.20}$&
               $\quad$$\Xi^{0}_{c}\rightarrow\Lambda^{0}\pi^{0}K^{0}_{L}$&
               $\quad$$(5.19\pm1.39)\times10^{-3}$&
               $\quad$$0.20\pm0.50$\\

               $\Xi^{+}_{c}\rightarrow\Xi^{0}K^{+}K^{0}_{L}$&
               $\quad$$(9.82\pm4.76)\times10^{-4}$&
               $\quad$$0.96\pm0.02$&
               $\quad$$\Xi^{0}_{c}\rightarrow\Sigma^{0}\eta^{0}K^{0}_{S}$&
               $\quad$$(4.61\pm0.78)\times10^{-4}$&
               $\quad$$-0.70\pm0.28$\\

               $\Xi^{+}_{c}\rightarrow\Lambda^{0}\pi^{+}K^{0}_{S}$&
               $\quad$$(4.02\pm3.32)\times10^{-3}$&
               $\quad$$-0.70^{+0.70}_{-0.30}$&
               $\quad$$\Xi^{0}_{c}\rightarrow\Sigma^{0}\eta^{0}K^{0}_{L}$&
               $\quad$$(5.01\pm0.83)\times10^{-4}$&
               $\quad$$-0.65\pm0.29$\\

               $\Xi^{+}_{c}\rightarrow\Lambda^{0}\pi^{+}K^{0}_{L}$&
               $\quad$$(2.32^{+2.71}_{-2.32})\times10^{-3}$&
               $\quad$$-0.30^{+0.90}_{-0.70}$&
               $\quad$$\quad$&
               $\quad$$\quad$&
               $\quad$$\quad$\\

               $\Xi^{+}_{c}\rightarrow pK^{0}_{S}K^{0}_{S}$&
               $\quad$$(1.31\pm0.35)\times10^{-2}$&
               $\quad$$-0.65^{+0.70}_{-0.35}$&
               $\quad$$\quad$&
               $\quad$$\quad$&
               $\quad$$\quad$\\

               $\Xi^{+}_{c}\rightarrow pK^{0}_{S}K^{0}_{L}$&
               $\quad$$(1.23\pm0.34)\times10^{-2}$&
               $\quad$$-0.67^{+0.70}_{-0.33}$&
               $\quad$$\quad$&
               $\quad$$\quad$&
               $\quad$$\quad$\\

               $\Xi^{+}_{c}\rightarrow pK^{0}_{L}K^{0}_{L}$&
               $\quad$$(1.16\pm0.35)\times10^{-2}$&
               $\quad$$-0.69^{+0.60}_{-0.31}$&
               $\quad$$\quad$&
               $\quad$$\quad$&
               $\quad$$\quad$\\
\hline
\end{tabular}

    \label{tab:Numerical result 7}
\end{table}

\section{Conclusions}\label{Discussions and conclusions}
In our study of the antitriplet charmed baryon three-body weak decay of $\mathbf{B}_{c}\rightarrow\mathbf{B}_{n}PP^{\prime}$ within the framework of the $SU(3)_{F}$ flavor symmetry, we have incorporated contributions from both $H(\bar{\mathbf{6}})$ and $H(\mathbf{15})$ to decompose the decays into 16 real amplitudes. Through a minimum $\chi^2$ fit to 28 updated experimental data points, we have achieved a new fit with $\chi^2/d.o.f=1.5$. This significantly improved the fitting results, affirming the validity of $SU(3)_{F}$ in charmed baryon three-body decays. Based on the branch ratio fitting results, we have analyzed potential sources of errors, including the remaining resonant and excited state contributions in the experimental data, the contributions of $a_{9}$ and $a_{10}$ in $H(15)$, and the P-wave contribution in the final states. We have given one of those predictions of $\mathcal{B}\,(\Lambda^{+}_{c}\rightarrow(\Xi(1690)^{0}\rightarrow\Sigma^{+}K^{-})\,K^{+})=(1.5\pm0.4)\times10^{-3}$, which needs to be further explored by experiments. Meanwhile, the possible $SU(3)_{F}$ symmetry breaking effect in processes, such as $\Xi^{+}_{c}\rightarrow p\pi^{+}K^{-}$ and $\Lambda^{+}_{c}\rightarrow \Sigma^{+}\pi^{-}K^{+}$, is notable in theoretical research. We have found that in the chiral limit, 
 the  $\langle\alpha\rangle$ of $\Lambda^{+}_{c}\rightarrow\Xi^{0}\pi^{0}K^{+}$ and $\Lambda^{+}_{c}\rightarrow\Xi^{0}K^{+}\eta^{0}$  is negative, which enables us to fix 
 $Z_2$ ambiguity in the $SU(3)_{F}$ fit. We have also updated decays involving $K^{0}_{S}$ and $K^{0}_{L}$. In particular, we find that three decay channels of $\Lambda^{+}_{c}\rightarrow n\pi^{+}\pi^{0}$, $\Lambda^{+}_{c}\rightarrow \Lambda^{0}K^{+}K^{0}$ and $\Xi^{+}_{c}\rightarrow \Lambda^{0}\pi^{+}\pi^{0}$ are only contributed by $H(\mathbf{15})$, prompting the further theoretical analysis in classification. Lastly, measurements with large errors of channels $\Xi^{+}_{c}\rightarrow\Sigma^{+}\pi^{+}\pi^{-}$, $\Xi^{+}_{c}\rightarrow\Xi^{0}\pi^{+}\pi^{0}$, $\Xi^{+}_{c}\rightarrow\Sigma^{-}\pi^{+}\pi^{+}$ and $\Xi^{+}_{c}\rightarrow\Sigma^{+}K^{+}K^{-}$ should be remeasured for higher precision.

\section*{Acknowledgement}
We would like to express our sincere appreciation to Dr. Jiabao Zhang for his valuable insights during the development of this work. This work is supported in part by the National Key Research and Development Program of China under Grant No. 2020YFC2201501 and  the National Natural Science Foundation of China (NSFC) under Grant No. 12347103 and 12205063.

\newpage
\appendix
\section{A-amplitude of $\mathbf{B}_{c}\rightarrow\mathbf{B}_{n}PP^{\prime}$}\label{appendix}

\begin{table}[H]
    \centering
    \caption{A-amplitudes of $\Lambda^{+}_{c}\rightarrow\mathbf{B}_{n}PP^{\prime}$}
    
\renewcommand{\arraystretch}{1}
\footnotesize
\begin{tabular}{llllll}
\hline
CF mode$\qquad$ &
  $Ac^{-2}_{c}$ &
  $\quad$CF mode$\qquad$ &
  $Ac^{-2}_{c}$
  \\ \hline
$\Lambda^{0}\pi^{+}\eta^{0}$ &
  $-\frac{2}{3}(a_{2}-a_{3}+a_{5}+3a_{6}-a_{7}+a_{8})$ &
  $\quad$$p\pi^{0}\bar{K}^{0}$ &
   $-\frac{1}{\sqrt{2}}(2a_{3}+2a_{4}-a_{7}+a_{8})$ 
  \\
$\Lambda^{0}K^{+}\bar{K}^{0}$ &
  $-\sqrt{\frac{2}{3}}(a_{2}-a_{3}+a_{5}-a_{7}-a_{8})$ &
  $\quad$$p\bar{K}^{0}\eta^{0}$ &
   $\frac{1}{\sqrt{6}}(-2a_{3}+2a_{4}+a_{7}-a_{8})$ 
  \\
$\Sigma^{0}\pi^{+}\pi^{0}$ &
  $-2a_{4}-2a_{6}$ &
  $\quad$$p\pi^{+}K^{-}$ &
  $2a_{3}-2a_{6}-a_{7}-a_{8}$ 
  \\
$\Sigma^{0}\pi^{+}\eta^{0}$ &
   $\frac{2}{\sqrt{3}}(a_{2}+a_{3}+a_{4}+a_{5})$ &
   $\quad$$n\pi^{+}\bar{K}^{0}$ &
  $-2a_{4}-2a_{6}-2a_{8}$ 
   \\
$\Sigma^{0}K^{+}\bar{K}^{0}$ &
  $\sqrt{2}(a_{2}+a_{3}+a_{5})$ &
  $\quad$$\Sigma^{+}\pi^{0}\pi^{0}$ &
   $4a_{1}+2a_{2}+2a_{3}+2a_{4}-2a_{5}$ 
   \\
$\Sigma^{-}\pi^{+}\pi^{+}$ &
  $-4a_{4}-4a_{6}$ &
  $\quad$$\Sigma^{+}\pi^{0}\eta^{0}$ &
   $-\frac{2}{\sqrt{3}}(a_{2}+a_{3}+a_{4}+a_{5})$ 
  \\
$\Xi^{-}\pi^{+}K^{+}$ &
  $-2a_{6}$ &
  $\quad$$\Sigma^{+}\eta^{0}\eta^{0}$ &
   $\frac{2}{3}(6a_{1}+a_{2}+a_{3}+a_{4}-a_{5})$ 
   \\
$\Xi^{0}\pi^{0}K^{+}$ &
  $-\sqrt{2}a_{5}$ &
  $\quad$$\Sigma^{+}\pi^{+}\pi^{-}$ &
   $4a_{1}+2a_{2}+2a_{3}-2a_{5}-2a_{6}$ 
   \\
$\Xi^{0}K^{+}\eta^{0}$ &
  $\sqrt{\frac{2}{3}}a_{5}$ &
  $\quad$$\Sigma^{+}K^{+}K^{-}$ &
  $4a_{1}-2a_{5}$ 
  \\
$\Xi^{0}\pi^{+}K^{0}$ &
  $-2a_{5}-2a_{6}$ &
  $\quad$$\Sigma^{+}K^{0}\bar{K}^{0}$ &
  $4a_{1}+2a_{2}+2a_{3}$ 
  \\
\hline
\hline
CS mode$\qquad$ &
  $Ac^{-1}_{c}s^{-1}_{c}$ &
  $\quad$CS mode$\qquad$ &
  $Ac^{-1}_{c}s^{-1}_{c}$ 
\\ 
\hline
  $\Lambda^{0}\pi^{0}K^{+}$ &
   $\frac{1}{\sqrt{3}}(-a_{2}+a_{3}+2a_{5}+a_{7}+a_{8})$ &
  $\quad$$p\eta^{0}\eta^{0}$ &
   \makecell[l]{$\frac{1}{3}(12a_{1}+2a_{2}+8a_{3}$\\$-4a_{4}-2a_{5}-3a_{7}+3a_{8})$}
  \\
  $\Lambda^{0}K^{+}\eta^{0}$ &
  $-\frac{1}{3}(-a_{2}+a_{3}+2a_{5}+6a_{6}+a_{7}+5a_{8})$ &
  $\quad$$p\pi^{+}\pi^{-}$ &
   $4a_{1}+2a_{2}-2a_{5}+a_{7}+a_{8}$
  \\
  $\Lambda^{0}\pi^{+}K^{0}$ &
   $-\sqrt{\frac{2}{3}}(a_{2}-a_{3}-2a_{5}-a_{7}+a_{8})$ &
  $\quad$$pK^{+}K^{-}$ &
   $4a_{1}+2a_{3}-2a_{5}-2a_{6}-a_{7}-a_{8}$
  \\
  $\Sigma^{0}\pi^{0}K^{+}$ &
   $a_{2}+a_{3}-2a_{4}-2a_{6}$ &
  $\quad$$pK^{0}\bar{K}^{0}$ &
  $4a_{1}+2a_{2}+2a_{3}+2a_{4}$
   \\
  $\Sigma^{0}K^{+}\eta^{0}$ &
   $\frac{1}{\sqrt{3}}(-a_{2}-a_{3}+2a_{4})$ &
  $\quad$$n\pi^{+}\pi^{0}$ &
  $-\sqrt{2}a_{8}$
   \\
  $\Sigma^{0}\pi^{+}K^{0}$ &
  $\sqrt{2}(a_{2}+a_{3}+a_{4})$ &
  $\quad$$n\pi^{+}\eta^{0}$ &
   $\sqrt{\frac{2}{3}}(-2a_{2}+2a_{4}-2a_{5}+a_{7}+2a_{8})$
  \\
  $\Sigma^{-}\pi^{+}K^{+}$ &
  $-4a_{4}-2a_{6}$ &
  $\quad$$nK^{+}\bar{K}^{0}$ &
   $-2a_{2}-2a_{4}-2a_{5}-2a_{6}+a_{7}-a_{8}$
   \\
  $\Xi^{-}K^{+}K^{+}$ &
  $-4a_{6}$ &
  $\quad$$\Sigma^{+}\pi^{0}K^{0}$ &
  $-\sqrt{2}(a_{2}+a_{3}+2a_{4})$
   \\
  $\Xi^{0}K^{+}K^{0}$ &
  $-2a_{6}$ &
  $\quad$$\Sigma^{+}K^{0}\eta^{0}$ &
  $\sqrt{\frac{2}{3}}(-a_{2}-a_{3}+2a_{4})$
  \\
  $p\pi^{0}\pi^{0}$ &
   $4a_{1}+2a_{2}-2a_{5}+a_{7}-a_{8}$ &
  $\quad$$\Sigma^{+}\pi^{-}K^{+}$ &
  $2a_{2}+2a_{3}-2a_{6}$
  \\
  $p\pi^{0}\eta^{0}$ &
   $\frac{1}{\sqrt{3}}(-2a_{2}+2a_{4}-2a_{5}-a_{7}+a_{8})$ &
  $\quad$ &
  $\quad$
  \\
\hline
\hline
  DCS mode$\qquad$ &
  $As^{-2}_{c}$ &
  $\quad$DCS mode$\qquad$ &
  $As^{-2}_{c}$ 
  \\ 
  \hline
  $\Lambda^{0}K^{+}K^{0}$ &
  $-\sqrt{\frac{8}{3}}a_{8}$ &
  $\quad$$p\pi^{-}K^{+}$ &
  $2a_{2}+a_{7}+a_{8}$
  \\
  $\Sigma^{0}K^{+}K^{0}$ &
  $2\sqrt{2}a_{4}$ &
  $\quad$$n\pi^{0}K^{+}$ &
   $-\frac{1}{\sqrt{2}}(2a_{2}-a_{7}+a_{8})$
  \\
  $\Sigma^{-}K^{+}K^{+}$ &
  $-4a_{4}$ &
  $\quad$$nK^{+}\eta^{0}$ &
   $\frac{1}{\sqrt{6}}(2a_{2}+4a_{4}-a_{7}+a_{8})$
  \\
  $p\pi^{0}K^{0}$ &
   $\frac{1}{\sqrt{2}}(-2a_{2}-a_{7}+a_{8})$ &
   $\quad$$n\pi^{+}K^{0}$ &
  $-2a_{2}+a_{7}+a_{8}$ 
   \\
  $pK^{0}\eta^{0}$ &
   $-\frac{1}{\sqrt{6}}(2a_{2}+4a_{4}+a_{7}-a_{8})$ &
  $\quad$$\Sigma^{+}K^{0}K^{0}$ &
  $4a_{4}$ 
   \\
  \hline
\end{tabular}

    \label{tab:A-amplitudes1}
\end{table}

\begin{table}[H]
    \centering
    \caption{A-amplitudes of $\Xi^{+}_{c}\rightarrow\mathbf{B}_{n}PP^{\prime}$}
    
\renewcommand{\arraystretch}{1}
\footnotesize
\begin{tabular}{llllll}
\hline
CF mode$\qquad$ &
  $Ac^{-2}_{c}$ &
  $\quad$CF mode$\qquad$ &
  $Ac^{-2}_{c}$
 \\ 
 \hline
$\Lambda^{0}\pi^{+}\bar{K}^{0}$ &
  $\sqrt{\frac{2}{3}}(3a_4+a_{8})$ &
  $\quad$$\Xi^{0}K^{+}\bar{K}^{0}$ &
  $-2a_{2}+a_{7}+a_{8}$
  \\
$\Sigma^{0}\pi^{+}\bar{K}^{0}$ &
  $\sqrt{2}(a_{4}-a_{8})$ &
 $\quad$$p\bar{K}^{0}\bar{K}^{0}$ &
  $4a_{4}$
  \\
$\Xi^{-}\pi^{+}\pi^{+}$ &
  $-4a_{4}$ &
  $\quad$$\Sigma^{+}\pi^{0}\bar{K}^{0}$ &
  $\frac{1}{\sqrt{2}}(-2a_{2}-2a_{4}-a_{7}+a_{8})$
  \\
$\Xi^{0}\pi^{+}\pi^{0}$ &
  $\sqrt{2}a_{4}$ &
  $\quad$$\Sigma^{+}\bar{K}^{0}\eta^{0}$ &
  $\frac{1}{\sqrt{6}}(-2a_{2}+2a_{4}-a_{7}+a_{8})$
  \\
$\Xi^{0}\pi^{+}\eta^{0}$ &
  $-\sqrt{\frac{2}{3}}(2a_{2}+a_{4}-a_{7}+a_{8})$ &
  $\quad$$\Sigma^{+}\pi^{+}K^{-}$ &
  $2a_{2}+a_{7}+a_{8}$
  \\
  \hline
  \hline
CS mode$\qquad$ &
  $Ac^{-1}_{c}s^{-1}_{c}$ &
  $\quad$CS mode$\qquad$ &
  $Ac^{-1}_{c}s^{-1}_{c}$ 
  \\ \hline
$\Lambda^{0}\pi^{+}\pi^{0}$ &
  $\frac{1}{\sqrt{3}}a_{8}$ &
  $\quad$$p\pi^{0}\bar{K}^{0}$ &
  $-\sqrt{2}(a_{2}+a_{3})$ 
  \\
$\Lambda^{0}\pi^{+}\eta^{0}$ &
 \makecell[l]{$-\frac{1}{3}(-4a_{2}-2a_{3}+6a_{4}$\\$+2a_{5}+6a_{6}+a_{7}+2a_{8})$} &
  $\quad$$p\bar{K}^{0}\eta^{0}$ &
  $-\sqrt{\frac{2}{3}}(a_{2}+a_{3}+4a_{4})$
  \\
$\Lambda^{0}K^{+}\bar{K}^{0}$ &
 $\frac{1}{\sqrt{6}}(4a_{2}+2a_{3}+6a_{4}-2a_{5}-a_{7}+a_{8})$ &
  $\quad$$p\pi^{+}K^{-}$ &
  $2a_{2}+2a_{3}-2a_{6}$ 
  \\
$\Sigma^{0}\pi^{+}\pi^{0}$ &
  $-2a_{6}-a_{8}$ &
  $\quad$$n\pi^{+}\bar{K}^{0}$ &
  $-2a_{6}$ 
  \\
$\Sigma^{0}\pi^{+}\eta^{0}$ &
  $\frac{1}{\sqrt{3}}(2a_{3}-2a_{4}+2a_{5}+a_{7}+2a_{8})$ &
  $\quad$$\Sigma^{+}\pi^{0}\pi^{0}$ &
  $4a_{1}+2a_{3}-2a_{5}-a_{7}+a_{8}$ 
  \\
$\Sigma^{0}K^{+}\bar{K}^{0}$ &
  $\frac{1}{\sqrt{2}}(2a_{3}+2a_{4}+2a_{5}+a_{7}-a_{8})$ &
  $\quad$$\Sigma^{+}\pi^{0}\eta^{0}$ &
  $\frac{1}{\sqrt{3}}(-2a_{3}+2a_{4}-2a_{5}+a_{7}-a_{8})$ 
  \\
$\Sigma^{-}\pi^{+}\pi^{+}$ &
  $-4a_{6}$ &
  $\quad$$\Sigma^{+}\eta^{0}\eta^{0}$ &
 \makecell[l]{$\frac{1}{3}(12a_{1}+8a_{2}+2a_{3}$\\$-4a_{4}-2a_{5}+3a_{7}-3a_{8})$} 
  \\
$\Xi^{-}\pi^{+}K^{+}$ &
  $-4a_{4}-2a_{6}$ &
  $\quad$$\Sigma^{+}\pi^{+}\pi^{-}$ &
  $4a_{1}+2a_{3}-2a_{5}-2a_{6}-a_{7}-a_{8}$ 
  \\
$\Xi^{0}\pi^{0}K^{+}$ &
  $\frac{1}{\sqrt{2}}(-2a_{2}+2a_{4}-2a_{5}+a_{7}+a_{8})$ &
  $\quad$$\Sigma^{+}K^{+}K^{-}$ &
  $4a_{1}+2a_{2}-2a_{5}+a_{7}+a_{8}$ 
  \\
$\Xi^{0}K^{+}\eta^{0}$ &
  $\frac{1}{\sqrt{6}}(2a_{2}-2a_{4}+2a_{5}-a_{7}-5a_{8})$ &
  $\quad$$\Sigma^{+}K^{0}\bar{K}^{0}$ &
  $4a_{1}+2a_{2}+2a_{3}+2a_{4}$ 
  \\
$\Xi^{0}\pi^{+}K^{0}$ &
  $-2a_{2}-2a_{4}-2a_{5}-2a_{6}+a_{7}-a_{8}$ &
  $\quad$ &
  $\quad$
  \\
  \hline
  \hline
DCS mode$\qquad$ &
  $As^{-2}_{c}$ &
  $\quad$DCS mode$\qquad$ &
  $As^{-2}_{c}$
  \\ \hline
$\Lambda^{0}\pi^{0}K^{+}$ &
  $\frac{1}{2\sqrt{3}}(4a_{2}+2a_{3}+4a_{5}-a_{7}+a_{8})$ &
  $\quad$$p\pi^{0}\eta^{0}$ &
  $-\frac{2}{\sqrt{3}}a_{5}$
  \\
$\Lambda^{0}K^{+}\eta^{0}$ &
 \makecell[l]{$-\frac{1}{6}(4a_{2}+2a_{3}+12a_{4}$\\$+4a_{5}+12a_{6}-a_{7}+a_{8})$} &
  $\quad$$p\eta^{0}\eta^{0}$ &
  $\frac{2}{3}(6a_{1}+4a_{2}+4a_{3}+4a_{4}-a_{5})$
  \\
$\Lambda^{0}\pi^{+}K^{0}$ &
  $\frac{1}{\sqrt{6}}(4a_{2}+2a_{3}+4a_{5}-a_{7}-a_{8})$ &
  $\quad$$p\pi^{+}\pi^{-}$ &
  $4a_{1}-2a_{5}$ 
  \\
$\Sigma^{0}\pi^{0}K^{+}$ &
  $\frac{1}{2}(2a_{3}-4a_{6}+a_{7}-a_{8})$ &
  $\quad$$pK^{+}K^{-}$ &
  $4a_{1}+2a_{2}+2a_{3}-2a_{5}-2a_{6}$ 
  \\
$\Sigma^{0}K^{+}\eta^{0}$ &
  $-\frac{1}{2\sqrt{3}}(2a_{3}+4a_{4}+a_{7}-a_{8})$ &
  $\quad$$pK^{0}\bar{K}^{0}$ &
  $4a_{1}+2a_{2}+2a_{3}$ 

  \\
$\Sigma^{0}\pi^{+}K^{0}$ &
  $\frac{1}{\sqrt{2}}(2a_{3}+a_{7}+a_{8})$ &
  $\quad$$n\pi^{+}\eta^{0}$ &
  $-\sqrt{\frac{8}{3}}a_{5}$ 

  \\
$\Sigma^{-}\pi^{+}K^{+}$ &
  $-2a_{6}$ &
  $\quad$$nK^{+}\bar{K}^{0}$ &
  $-2a_{5}-2a_{6}$ 

  \\
$\Xi^{-}K^{+}K^{+}$ &
  $-4a_{4}-4a_{6}$ &
  $\quad$$\Sigma^{+}\pi^{0}K^{0}$ &
  $\frac{1}{\sqrt{2}}(-2a_{3}+a_{7}-a_{8})$ 

  \\
$\Xi^{0}K^{+}K^{0}$ &
  $-2a_{4}-2a_{6}-2a_{8}$ &
  $\quad$$\Sigma^{+}K^{0}\eta^{0}$&
  $-\frac{1}{\sqrt{6}}(2a_{3}+4a_{4}-a_{7}+a_{8})$ 

  \\
$p\pi^{0}\pi^{0}$ &
  $4a_{1}-2a_{5}$ &
  $\quad$$\Sigma^{+}\pi^{-}K^{+}$ &
  $2a_{3}-2a_{6}-a_{7}-a_{8}$ 
  \\
  \hline
\end{tabular}

    \label{tab:A-amplitudes2}
\end{table}

\begin{table}[H]
    \centering
    \caption{A-amplitudes of CF mode and CS mode $\Xi^{0}_{c}\rightarrow\mathbf{B}_{n}PP^{\prime}$}
    
\renewcommand{\arraystretch}{1}
\footnotesize
\begin{tabular}{llllll}
\hline
CF mode$\qquad$ &
  $Ac^{-2}_{c}$ &
  $\quad$CF mode$\qquad$ &
  $Ac^{-2}_{c}$ 
  \\ \hline
$\Lambda^{0}\pi^{0}\bar{K}^{0}$&
  $-\frac{1}{2\sqrt{3}}(2a_{2}+4a_{3}+6a_{4}+2a_{5}-a_{7}+a_{8})$&
  $\quad$$\Xi^{0}\pi^{0}\eta^{0}$ &
  $\frac{2}{\sqrt{3}}(a_{2}+a_{3}+a_{4})$ 

  \\
$\Lambda^{0}\bar{K}^{0}\eta^{0}$ &
  \makecell[l]{$-\frac{1}{6}(2a_{2}+4a_{3}-6a_{4}$\\$+2a_{5}-12a_{6}-a_{7}+a_{8})$} &
  $\quad$$\Xi^{0}\eta^{0}\eta^{0}$ &
  $-\frac{2}{3}(6 a_{1}+ a_{2}+ a_{3}+ a_{4}-4 a_{5})$ 

  \\
$\Lambda^{0}\pi^{+}{K}^{-}$ &
  $\frac{1}{\sqrt{6}}(2 a_{2}+4 a_{3}+2 a_{5}-a_{7}-a_{8})$ &
  $\quad$$\Xi^{0}\pi^{+}\pi^{-}$&
  $-4 a_{1}-2a_{2}-2a_{3}$ 

  \\
$\Sigma^{0}\pi^{0}\bar{K}^{0}$ &
  $\frac{1}{2}(2 a_{2}+2 a_{4}+2 a_{5}+4 a_{6}+a_{7}-a_{8})$ &
  $\quad$$\Xi^{0}K^{+}K^{-}$ &
  $-4a_{1}+2a_{5}$ 

  \\
$\Sigma^{0}\bar{K}^{0}\eta^{0}$ &
  $\frac{1}{2 \sqrt{3}}(2 a_{2}-2 a_{4}+2 a_{5}+a_{7}-a_{8})$ &
  $\quad$$\Xi^{0}K^{0}\bar{K}^{0}$ &
  $-4 a_{1}-2a_{2}-2a_{3}+2a_{5}+2a_{6}$ 

  \\
$\Sigma^{0}\pi^{+}K^{-}$ &
  $-\frac{1}{\sqrt{2}}(2 a_{2}+2 a_{5}+a_{7}+a_{8})$ &
  $\quad$$pK^{-}\bar{K}^{0}$ &
  $2 a_{6} $ 

\\
$\Sigma^{-}\pi^{+}\bar{K}^{0}$ &
  $2 a_{4}+2 a_{6}-2 a_{8}$ &
  $\quad$$n\bar{K}^{0}\bar{K}^{0}$ &
  $4 a_{4}+4 a_{6}$ 

  \\
$\Xi^{-}\pi^{+}\pi^{0}$ &
  $\sqrt{2}a_{4}$ &
  $\quad$$\Sigma^{+}\pi^{0}K^{-}$ &
  $\sqrt{2} a_{5}$ 

  \\
$\Xi^{-}\pi^{+}\eta^{0}$ &
  $-\sqrt{\frac{2}{3}}(2a_{3}+a_{4}+a_{7}-a_{8})$ &
  $\quad$$\Sigma^{+}K^{-}\eta^{0}$ &
  $-\sqrt{\frac{2}{3}}a_{5}$ 

  \\
$\Xi^{-}K^{+}\bar{K}^{0}$ &
  $-2 a_{3}+2 a_{6}-a_{7}-a_{8}$ &
  $\quad$$\Sigma^{+}\pi^{-}\bar{K}^{0}$ &
  $2 a_{5}+2a_{6}$ 
  \\
$\Xi^{0}\pi^{0}\pi^{0}$ &
  $-4 a_{1}-2 a_{2}-2 a_{3}-2 a_{4}$ &
  $\quad$&
  $\quad$
  \\ \hline
  \hline
CS mode$\qquad$ &
  $Ac^{-1}_{c}s^{-1}_{c}$ &
  $\quad$CS mode$\qquad$ &
  $Ac^{-1}_{c}s^{-1}_{c}$ 
  \\ \hline
$\Lambda^{0}\pi^{0}\pi^{0}$ &
 $\frac{1}{\sqrt{6}}(12 a_{1}+4 a_{2}+2 a_{3}-2 a_{5}+a_{7}-a_{8})$&
  $\quad$$\Xi^{-}\pi^{0}K^{+}$ &
  $-\frac{1}{\sqrt{2}}(2 a_{3}-2 a_{4}-2 a_{6}+a_{7}+a_{8})$ 

  \\
$\Lambda^{0}\pi^{0}\eta^{0}$ &
  \makecell[l]{$\frac{1}{3\sqrt{2}}(-4 a_{2}-2 a_{3}+6 a_{4}$\\$+2 a_{5}+6 a_{6}-a_{7}+a_{8})$} &
  $\quad$$\Xi^{-}K^{+}\eta^{0}$ &
  $\frac{1}{\sqrt{6}} (2a_{3}-2 a_{4}-6 a_{6}+a_{7}+5 a_{8})$ 

  \\
$\Lambda^{0}\eta^{0}\eta^{0}$ &
  \makecell[l]{$\frac{1}{\sqrt{6}}(12 a_{1}+4 a_{2}+6 a_{3}-4 a_{4}$\\$-6 a_{5}-12 a_{6}-a_{7}+a_{8})$} &
  $\quad$$\Xi^{-}\pi^{+}K^{0}$ &
  $-2 a_{3}-2 a_{4}-a_{7}+a_{8}$

  \\
$\Lambda^{0}\pi^{+}\pi^{-}$ &
  $\frac{1}{\sqrt{6}}(12 a_{1}+4 a_{2}+2 a_{3}-2 a_{5}+a_{7}+a_{8})$
&
  $\quad$$\Xi^{0}\pi^{0}K^{0}$ &
  $\sqrt{2}( a_{2}+a_{3}+2 a_{4}+a_{5}+a_{6})$ 

  \\
$\Lambda^{0}K^{+}K^{-}$ &
  $\frac{1}{\sqrt{6}}(12 a_{1}+2 a_{2}+4 a_{3}-4 a_{5}-a_{7}-a_{8})$
&
  $\quad$$\Xi^{0}K^{0}\eta^{0}$ &
  $\sqrt{\frac{2}{3}}( a_{2}+a_{3}-2 a_{4}+a_{5}-3 a_{6})$ 
  \\
$\Lambda^{0}K^{0}\bar{K}^{0}$ &
 $\sqrt{6}(2 a_{1}+ a_{2}+ a_{3}+ a_{4}- a_{5})$ &
  $\quad$$\Xi^{0}\pi^{-}K^{+}$ &
  $-2 a_{2}-2a_{3}-2a_{5}$ 

\\
$\Sigma^{0}\pi^{0}\pi^{0}$ &
  $-\frac{1}{\sqrt{2}}(4 a_{1}+2 a_{3}-2 a_{5}-4 a_{6}-a_{7}+a_{8})$ 
&
  $\quad$$p\pi^{0}K^{-}$ &
  $\sqrt{2} (a_{5}+a_{6})$ 

  \\
$\Sigma^{0}\pi^{0}\eta^{0}$ &
  $\frac{1}{\sqrt{6}}(2 a_{3}-2 a_{4}-2 a_{5}-6 a_{6}-a_{7}+a_{8})$
&
  $\quad$$pK^{-}\eta^{0}$ &
  $-\sqrt{\frac{2}{3}}(a_{5}+3 a_{6})$ 

  \\
$\Sigma^{0}\eta^{0}\eta^{0}$ &
  \makecell[l]{$-\frac{1}{3 \sqrt{2}}(12 a_{1}+8 a_{2}+2 a_{3}$\\$-4 a_{4}-2 a_{5}+3 a_{7}-3 a_{8})$} &
  $\quad$$p\pi^{-}\bar{K}^{0}$ &
  $2a_{5}$ 

  \\
$\Sigma^{0}\pi^{+}\pi^{-}$ &
  $\frac{1}{\sqrt{2}}(-4 a_{1}-2 a_{3}+2 a_{5}+a_{7}+a_{8})$ 
&
  $\quad$$n\pi^{0}\bar{K}^{0}$ &
  $-\sqrt{2}(a_{2}+a_{3}+a_{5}-a_{6})$ 
  \\
$\Sigma^{0}K^{+}K^{-}$ &
  $-\frac{1}{\sqrt{2}}(4 a_{1}+2 a_{2}+a_{7}+a_{8})$ 
&
  $\quad$$n\bar{K}^{0}\eta^{0}$ &
  $-\sqrt{\frac{2}{3}}( a_{2}+a_{3}+4 a_{4}+a_{5}+3 a_{6})$ 

  \\
$\Sigma^{0}K^{0}\bar{K}^{0}$ &
  $-\sqrt{2}(2 a_{1}+ a_{2}+ a_{3}+ a_{4}- a_{5})$ &
  $\quad$$n\pi^{+}K^{-}$ &
  $2 a_{2}+2a_{3}+2a_{5}$ 

  \\
$\Sigma^{-}\pi^{+}\pi^{0}$ &
  $\sqrt{2} (a_{6}-a_{8})$ 
&
  $\quad$$\Sigma^{+}\pi^{0}\pi^{-}$ &
  $\sqrt{2} a_{6}$ 

  \\
$\Sigma^{-}\pi^{+}\eta^{0}$ &
  $\sqrt{\frac{2}{3}}(2 a_{3}-2 a_{4}-3 a_{6}+a_{7}+2 a_{8})$ 
&
  $\quad$$\Sigma^{+}\pi^{-}\eta^{0}$ &
  $-\sqrt{\frac{2}{3}}(2a_{5}+3a_{6})$ 

  \\
$\Sigma^{-}K^{+}\bar{K}^{0}$ &
  $2 a_{3}+2 a_{4}+a_{7}-a_{8}$ &
  $\quad$$\Sigma^{+}K^{0}K^{-}$ &
  $-2a_{5}$ 
  \\
 \hline
\end{tabular}

    \label{tab:A-amplitudes3.1}
\end{table}

\begin{table}[H]
    \centering
    \caption{(Continued) A-amplitudes of DCS mode $\Xi^{0}_{c}\rightarrow\mathbf{B}_{n}PP^{\prime}$}
    
\renewcommand{\arraystretch}{1}
\footnotesize
\begin{tabular}{llllll}
\hline
DCS mode$\qquad$ &
  $As^{-2}_{c}$ &
  $\quad$DCS mode$\qquad$ &
  $As^{-2}_{c}$ &
  &
  \\ \hline
$\Lambda^{0}\pi^{0}K^{0}$ &
 $\frac{1}{2 \sqrt{3}}(-4 a_{2}-2 a_{3}-4 a_{5}-a_{7}+a_{8})$ 
&
  $\quad$$\Xi^{0}K^{0}K^{0}$ &
  $-4 a_{4}-4 a_{6}$ 

  \\
$\Lambda^{0}K^{0}\eta^{0}$ &
  \makecell[l]{$-\frac{1}{6}(4 a_{2}+2 a_{3}+12 a_{4}$\\$+4 a_{5}+12 a_{6}+a_{7}-a_{8})$} 
&
  $\quad$$p\pi^{-}\eta^{0}$ &
  $-\sqrt{\frac{8}{3}} a_{5} $ 

  \\
$\Lambda^{0}\pi^{-}K^{+}$ &
  $\frac{1}{\sqrt{6}}(4 a_{2}+2 a_{3}+4 a_{5}+a_{7}+a_{8})$ 
&
  $\quad$$pK^{0}K^{-}$ &
  $-2 a_{5}-2a_{6}$ 

  \\
$\Sigma^{0}\pi^{0}K^{0}$ &
  $\frac{1}{2}(2 a_{3}-4 a_{6}-a_{7}+a_{8})$ 
&
  $\quad$$n\pi^{0}\pi^{0}$ &
  $4a_{1}-2a_{5}$ 

  \\
$\Sigma^{0}K^{0}\eta^{0}$ &
  $\frac{1}{2 \sqrt{3}}(2 a_{3}+4 a_{4}-a_{7}+a_{8})$ &
  $\quad$$n\pi^{0}\eta^{0}$ &
  $\frac{2}{\sqrt{3}}a_{5}$ 
  \\
$\Sigma^{0}\pi^{-}K^{+}$ &
 $\frac{1}{\sqrt{2}}(-2 a_{3}+a_{7}+a_{8})$ 
&
  $\quad$$n\eta^{0}\eta^{0}$ &
  $\frac{2}{3} (6 a_{1}+4 a_{2}+4 a_{3}+4 a_{4}-a_{5})$ 

\\
$\Sigma^{-}\pi^{0}K^{+}$ &
  $\frac{1}{\sqrt{2}}(2 a_{3}+a_{7}-a_{8})$ 
&
  $\quad$$n\pi^{+}\pi^{-}$ &
  $4a_{1}-2a_{5} $ 

  \\
$\Sigma^{-}K^{+}\eta^{0}$ &
  $-\frac{1}{\sqrt{6}}(2 a_{3}+4 a_{4}+a_{7}-a_{8})$ 
&
  $\quad$$nK^{+}K^{-}$ &
  $4 a_{1}+2 a_{2}+2a_{3}$ 

  \\
$\Sigma^{-}\pi^{+}K^{0}$ &
  $2 a_{3}-2 a_{6}+a_{7}+a_{8}$ 
&
  $\quad$$nK^{0}\bar{K}^{0}$ &
  $4 a_{1}+2 a_{2}+2a_{3}-2a_{5}-2a_{6}$ 

  \\
$\Xi^{-}K^{+}\bar{K}^{0}$ &
  $-2 a_{4}-2 a_{6}+2 a_{8}$ &
  $\quad$$\Sigma^{+}\pi^{-}K^{0}$ &
  $-2 a_{6} $ 
  \\
  \hline
\end{tabular}

    \label{tab:A-amplitudes3.2}
\end{table}

\end{document}